\addcontentsline{toc}{section}{Preface}
\documentclass[12pt,amsmath,amssymb]{article}
\usepackage{graphicx}
\usepackage{color, soul}
\usepackage[toc]{appendix}
\usepackage{amsmath}
\usepackage{amsfonts}
\usepackage{dcolumn}
\usepackage{zref-totpages}
\input epsf
\usepackage{epigraph}
\usepackage[pdftex,
            pdfauthor={V. A. Yurov, A. V. Yurov},
            pdftitle={Quantum Cosmology Without Singularities: A New Approach},
            pdfsubject={},
            pdfkeywords={Many Interacting Worlds Interpretation, Cosmology, Quantum Gravity, cosmological singularities, zero universes},
            pdfproducer={Latex with hyperref},
            pdfcreator={pdflatex}]{hyperref}
\hypersetup{colorlinks=true, citecolor=blue, linkcolor=red}
\usepackage{enumerate}
\usepackage{scrextend}
\addtokomafont{labelinglabel}{\sffamily}
\newcommand{\beq}{\begin{equation}}
\newcommand{\beqn}{\begin{equation*}}
\newcommand{\enq}{\end{equation}}
\newcommand{\enqn}{\end{equation*}}
\newcommand{\bs}{\begin{subequations}}
\newcommand{\es}{\end{subequations}}
\newcommand{\eb}{{\rm e}}
\newcommand{\R}{{\mathbb R}}
\newcommand{\N}{{\mathbb N}}

\newcommand{\C}{{\mathbb C}}

\newcommand{\vphi}{\varphi}
\renewcommand{\eb}{\mathrm{e}}

\newtheorem{theorem}{\textsc{Theorem}}

\newtheorem{thm}{Theorem}

\newtheorem{condition}[thm]{Condition}

\newcommand{\p}{{\mathbf p}}

\def\({\left(}
\def\){\right)}

\newcommand{\footremember}[2]{%
    \footnote{#2}
    \newcounter{#1}
    \setcounter{#1}{\value{footnote}}%
}
\newcommand{\footrecall}[1]{%
    \footnotemark[\value{#1}]%
}

\begin{document}
\title{Quantum Cosmology Without Singularities: A New Approach}
\author{
Valerian A. Yurov\footremember{Kant}{Immanuel Kant Baltic Federal University, Institute of Physics, Mathematics and Informational Technology,
 Al.Nevsky St. 14, Kaliningrad, 236041, Russia} \footnote{vayt37@gmail.com} \and
Artyom V. Yurov \footrecall{Kant} \footnote{artyom\_yurov@mail.ru}
}


\date{\today}
\maketitle
\begin{abstract}
The article is dedicated to a discussion regarding the role of Barrow's ''Zero Universes'' in quantum cosmology. In particular, we demonstrate that if quantum gravity effects are modeled by the quantum potential method associated with the ''many interacting universes'' (MIU) model, then the mere presence of the universes with a zero scale factor (the ``Zero universes'') produces a veritably remarkable outcome: the classical cosmological singularities of Big Bang, Big Crunch and Big Rip all fail to arise. In other words, those universes that are considered ill-posed at the classical level may turn out to be a necessary and sought-after ingredient in a future internally consistent quantum theory of gravity. Finally, we argue that the MIU quantization method might shed light on a number of other cosmological mysteries; for example, it might account for a decoherence which preceded the eternal inflation, and elucidate how the quantum superposition of vacuum decays occurring at different places might give birth to actual bubble universes there. In addition, the new method might help explain why our universe was born in an extremely low-entropy initial state required to trigger the initial inflation.
\end{abstract}


\maketitle

\section*{Preface}
\addcontentsline{toc}{section}{Preface}

A reader might wonder: why start a scientific article with a Preface -- a mode of introduction seldomly used outside of anthologies, monographes and second editions? If the authors have something important to spill, why not pour it all into the Introduction, as is common and proper?.. The answer is rather simple. Just as this paper was receiving its final polish, we, the authors, have reached a chilling realization: in our quest of explaining all the methods and reasonings in as many details as possible (so that even an inattentive reader would not struggle too much following the article's logic), we actually overshoot by not providing a potential reader with a brief, straightforward explanation as to what this article actually is and isn't about. This realization was very troubling because, as everyone knows, a modern reader is all too eager to dump an unread article into an unreliable LLM chatbot, only to receive a rather questionable summary of a dubious value, and then regard the original article as ``banal and ''. So, we had no choice but to write this Preface!

First things first, let us begin by looking at what this article is about\footnote{Besides being about \ztotpages~pages long.}. It is dedicated to a quantization method developed by Hall, Deckert and Wiseman, known as the Many Interacting Worlds approach (MIW)  \cite{MIW}, with particular applications to gravity and cosmology, which we have called the Many Interacting Universes model (MIU). In fact, the article is an indirect sequel of the previously published paper \cite{YY} -- we say ``indirect'' because: (i) it employs a number of new ideas, and (ii) we took extra care to ensure that the article can be read with no previous knowledge of \cite{YY}.

The key results of the paper can be summed up as follows:

\begin{enumerate}
\item The quantization of cosmological solutions via the Hall, Deckert and Wiseman approach appears to be better suited for the expanding Friedman universes than the standard quantization methods, owing to the presence of cosmological horizons. The method becomes especially relevant for de Sitter (dS) universes, where the horizon is a key cosmological features -- in fact, for dS the MIW approach might be the only logical one (see Sec. \ref{Sec:MMIIUU}).

\item While an early Friedman universe is effectively ``smoothed over'' by inflation, on the later stages it might experience a direct manifestation of quantum effects on a scale of a cosmological horizon. It is important to note, that we are not referring here to a well known phenomena of quantum fluctuations being ``blown up'' by inflation and subsequently evolving into contemporary galaxies. What we are claiming instead is that the quantum effects can potentially influence a {\em dynamical expansion} of the universe, and might therefore be observable not just on an atomic scale, but on a cosmological scale (about $10^{28}$ cm) as well (see also \cite{YY}). In order to avoid a possible confusion, let us stress here that the quantum phenomena {\em do not} manifest themselves on the intermediate scales (between the micro and the mega), where a quantum decoherence reigns supreme.

\item At the core of our approach lies a model with a set (finite or countable) $\mathcal{D}$ of universes interacting via the Hall-Deckert-Wiseman quantum potential (see Appendix \ref{sec:Appendix} for its explanation, derivation, and relationship to a famous de Broglie-Bohm potential). As we shall soon see, by (i) adding to this set a Barrows's zero universe (a seemingly ill-defined object of zero scale factor \cite{Zero-Univ}), complimented with (ii) an additional empty universe with no fields of matter, and then (iii) by imposing a few simple physically sensible assumptions (see  Sec. \ref{Sec:MIU},  conditions 1-3), a most unusual Theorem can be proven -- that the Hall-Deckert-Wiseman quantization automatically eliminates the singularities with zero values of a scale factor (see Sec. \ref{Sec:no-singularity}, Theorem \ref{Theorem1}). For instance, a collapsing universe on an approach to a Big Crunch singularity experiences an effective rebound and undergos a new phase of expansion; for this no additional assumptions or fields of matter are required.

\item Additionally, a quantization under the aforementioned three conditions automatically gets rid of any Big Rip singularities (BR). In other words, any universe from $\mathcal{D}$ containing phantom fields with arbitrary equation of state (as long as $w<-1$) are effectively shielded from experiencing BR. This remarkable property is a purely kinematic phenomena and is a Corollary of the aforementioned Theorem (see Sec. \ref{Sec:Phantom}, Theorem \ref{Theorem2}).

\item Although on a first glance a Barrow universe with zero scale factor \cite{Zero-Univ} seems to be ill-posed, it is indeed an exact solution of Einstein equations. Furthermore, this solutions is merely a special case of the  Einstein quintessence model, which describes a closed stationary Einstein universe (see Sec. \ref{Sec:EQ}) -- and is itself a special case of a Tipler's Omega point solution. The fact that a zero universe is a full-fledged (albeit exotic) exact solution means that it has to be accounted for during the quantization procedure, be it via the functional integral method or by the Hall-Deckert-Wiseman approach. To put it in other words, an addition of the ``Barrow universe'' into a list of interacting universes is neither artificial nor a whim, but a natural and even essential step for a quantization of gravity.

\end{enumerate}

Again, we emphasize that these five statements are not hypothetical -- they are the results derived and (in case of the two theorems) proven in the current paper. However, we might add one more statement to this list, admittedly a more speculative one, but which nevertheless appears to be a logical conclusion of the Hall-Deckert-Wiseman formalism applied to an early universe:

\begin{enumerate}
\setcounter{enumi}{5}
\item The MIU method appears to provide interesting solutions for two fundamental problems of inflationary theory: for the decoherence conundrum \cite{Susk-B} and for the mystery of extremely low initial entropy, indispensable for jump-starting the process of inflation \cite{Entropy1}, \cite{Entropy2}.
\end{enumerate}

Again, we must warn the reader that we don't have the same rigorous proof for the last statement as we do for the previous five. However, as we discuss in Sec. \ref{Sec:discussion}, there are some evidences in favour of this statement.

Before we wrap the Preface up, in the spirit of completeness we would like to adumbrate the limitations of both the method and the results of this paper.

\begin{itemize}
\item  The paper is dedicated solely to singularities with zero values of scale factor, with an exception of Sec. \ref{Sec:Phantom}, which throws the Big Rip singularities into the mix. In order to achieve this we impose strict energy conditions upon the types of matter that fill the interacting universes. Thus, we are automatically dismissing possible singularities arising for finite non-zero values of scale factor, such as  the Big Freeze singularity \cite{3.1}, \cite{3.2},\cite{3.3}, the singularities localized on a brane \cite{3.4}, the Sudden Future singularity \cite{5.1}, \cite{5.2},\cite{5.3},  \cite{5.4}, \cite{5.5}, the Big Boost singularity \cite{BBtS}, and the Big Break singularity \cite{BBS1}, \cite{BBS2}\footnote{For the readers interested in classification of these singularities, we would like to refer them to the useful article \cite{NOT}.}. As a matter of fact, some of them -- like a Sudden singularity (with a finite value of the first derivative of scale factor) -- might still be included into our considerations. However, for the purposes of this article we are deliberately excluding {\em almost} anything that can be considered ``exotic'', restricting ourselves solely to the phenomena produced by the quantum potential  (\ref{U}). For that end, we are also abstaining from the sudden-like singularity from the \cite{SL} and from the unusual types of singularities that manifests themselves as finite jump discontinuities of cosmological parameters \cite{JC}. A detailed study of such late-time singularities, which significantly differ from the traditional Big Bang/Big Crunch, has been done in \cite{JOPAQuanSing} via the standard quantization approaches.

\item We have restricted ourselves to standard Einstein equations, without resorting to the models of modified gravity. This was done for a reason: in order to apply the MIU method of quantization to modified gravity models, the method itself must be modified, which lies far out of scope of this paper. An example of quantum cosmological models constructed via the traditional quantization recipes can be found in \cite{JOPAVasilev}.

\item On a cosmological scale the quantum effects manifests themselves as additional forces of repulsion, contributing to an accelerated expansion of the universe. So it would be natural to wonder, whether the quantum potential might actually explain the phenomenon of dark energy in observable universe. As far as we know, the answer is likely negative: the quantum interaction between the universes does not seem to be enough to explain observable acceleration of our universe -- a preliminary attempt reconciling the predictions of such a model with existing measurements on supernova apparent magnitude, Hubble parameter and baryon acoustic oscillations proved to be unsuccessful \cite{Astashenok}. On the other hand, the quantum potential might contribute to this acceleration, which implies that the dark energy might be a compound of both cosmological constant {\em and} of the quantum potential. However, verifying this hypothesis also lies out of scope of the current paper.
\end{itemize}

And now -- let's finally get on with the show!

\section{Introduction}\label{Sec:intro}
\subsection{An Overture for MIW}

When Andrei Linde famously quipped that ''quantum cosmology is conceptually one of the most difficult branches of theoretical physics'' \cite{Linde2005} it was the year 2005. Twenty years later, this quote has aged as a fine wine. One might argue that the difficulties starts with the very concept of ``quantum cosmology'', which sounds inherently catachrestic, as if it was a scientific Zen K\=oan; after all, we are dealing with an area of research purporting to describe the universe on a large scale (``cosmology'') whilst utilizing the small-scale (``quantum'') laws and principals. And yet, no one can deny the role, respect and reputation which quantum cosmology has amassed over the years. The triumphant rise of the theory of cosmological inflation in 1980-s has convinced even the most sceptical of physicists that the small scale quantum laws do have huge ramifications for the evolution of the universe at large. For instance, we now believe that the contemporary galaxies have germinated out of tiniest quantum fluctuations, instantaneously blown up out of proportions in the course of cosmological inflation \cite{MuhChib1981}, \cite{Hawk1982}, \cite{Star1982}. In other words, the cosmologists have resigned themselves to treating the very early stages in the evolution of the universe as a domain where the quantum gravity reign supreme. Admittedly, this view does restrict the arena of quantum cosmology to a supposedly extremely short time interval: one curious blip for the entire 14 billion years history of the universe. It is this discrepancy in time scale which leaves an unfortunately widespread impression that the entire field of quantum cosmology is rather ``esoteric'' -- if not to say ``superfluous''. And it is this very impression which we are going to confront in the current article.

The good place to start will be from a keystone of quantum cosmology: the Wheeler-DeWitt equation  (WDE) \cite{DeWitt1967}, \cite{Wheeler1968}.  WDE is essentially a Schr\"odinger equation defined over a super-space of all $(3+1)$-dimensional metrics; of particular interest here is a minisuperspace consisting of all possible Friedman metrics, which will be a subject of our scrutiny for the majority of this article (with an important exception of Section \ref{Sec:discussion}). Let us very briefly outline the key steps in the derivation of WDE, as they will prove to be important for our later discussion. One usually begins by defining the Friedman metrics for a closed universe, with the function $g_{00}=\mathcal{N}^2(t)$ being of particular significance. On the next step one writes down the standard Lagrangian; in the simplest case all fields of matter are treated as functions of a single minimally coupled scalar field $\phi$. The third step consists of integrating the Lagrangian w.r.t. the angular variables, thus producing the ``mechanical'' Lagrange function, which depends solely upon the variables $a(t)$, $\phi(t)$, $\mathcal{N}(t)$, and their derivatives ${\dot a}$, ${\dot\phi}$. Once the Lagrange function is in our grasp, we are free to define three canonical momenta, and then use the Legendre transform in order to construct the Hamilton function. At this step, two points should make themselves obvious: first, that that the function $\mathcal{N}$ will be a general multiple of the entire Hamilton function, and, secondly, that the canonical momentum for $\mathcal{N}$ is equal to zero (the latter is true since the Lagrange function does not depend upon the ``velocity'' ${\mathcal{\dot N}}$). These two facts, coupled with the Hamilton equations, lead us to an inevitable conclusion: {\em the Hamiltonian must be equal to zero}. Hence, after the quantization, when one replaces the canonical momenta with the corresponding operators of differentiation, a new equation will materialize: ${\mathcal H}\psi=0$, where $\psi=\psi(a,\phi)$ does not explicitly depend upon time. This new equation is, of course, nothing else but the sought after Wheeler-DeWitt equation. Again, let us emphasize: the {\em WDE is derived via the standard procedure of quantization}, when the dynamic variables within the classical Hamiltonian are effectively replaced by the operators (so that all the Poisson brackets morph into the commutators). The only distinctly cosmological steps in this process are related to the choices leading to the particular form of classical Hamiltonian, followed by the standard quantization techniques. And this is the key point where we are going to diverge from the WDE framework!

In this article we will make use of a different approach to the quantization procedure, first proposed in 2014 by Michael J. W. Hall, Dirk-Andre Deckert and Howard M. Wiseman in \cite{MIW}. They have demonstrated that the standard non-relativistic quantum mechanics arises as a continuous limit of a purely mechanical model, consisting of a large but finite ensemble $\mathcal{D}$ of classical ``worlds'' (they may be individual particles or entire systems of particles) interacting with each other via a special form of ``quantum potential'' -- so called because all the cumulative quantum effects and phenomena are but a product of this interaction. We have included the detailed explanation of the philosophy and derivations behind this ``Many Ineracting Worlds'' (MIW) approach in Appendix  \ref{sec:Appendix}, so here we will only mention key postulates that will be important later. According to the model, an evolution of each individual ``world'' is completely deterministic; the quantum mechanical probabilities are merely a by-product of the observer's ignorance about which world in the ensemble $\mathcal{D}$ is truly theirs. Fascinatingly, the MIW model correctly predicts and describes all quantum phenomena; an accuracy of this description grows rapidly along with the size $N$ of the ensemble, completely converging with the standard quantum field theory at the limit $N\to \infty$. Another noteworthy point is that, because there is no wave functions in the MIW, there is also no decoherence: the classical dynamics arises when and only when the quantum interaction becomes negligible.

In fact, this transition from quantum to classical description within MIW deserves a closer scrutiny. Let us consider two objects: a single baryon and a macroscopic object -- say, a tennis ball. Their behaviors are determined by two different theories: quantum mechanics for the former, classical Newtonian mechanics for the latter\footnote{The fastest tennis serve in the recorded history was at 263.4 km/h, performed by Sam Groth in 2012. This makes the fastest tennis balls move 4 million times slower then the speed of light: the tennis balls are strictly non-relativistic objects!}. Now, why would that be according to the MIW? The answer is pretty simple: a single baryon must by definition experience a force generated by the quantum potential \eqref{UNN} (see Sec. \ref{ssec:MIW} for more details), which is a result of interaction with the baryon's neighbours within $\mathcal{D}$. In a simplest one-dimensional case, a baryon from the ``world'' number $j$ will be most strongly affected by other baryons from the neighbouring ``worlds'' with numbers $j\pm 1$, whose contributions to the quantum potential will generally be rather large (otherwise the baryon will be completely classical). On the other hand, the tennis ball contains a huge amount of molecules, on the order of $N_{_A}\sim 10^{24}$ -- each one consisting of at least (!) 2 atoms. According to MIW, in order for the ball to act like a quantum object the majority of its constituent baryons must experience significant {\em codirectional} quantum force from their neighboring ``worlds'' -- otherwise the average cumulative force will be equal to zero! For example, if a given baryon experiences the required quantum force with probability $0.99999$ (a virtual certainty!), then for $N_{_A}$ baryons the said probability will be be reduced to $\exp\left(-10^{19}\right)$ -- a virtual impossibility! It is no wonder now why we can treat tennis balls so reliably as completely classical objects!

This little thought experiment can be generalized as the following statement: in MIW framework the objects with few degrees of freedom (such as elementary particles) must experience strong quantum interaction with their neighbours from ensemble $\mathcal{D}$, and will therefore act quantum mechanically. On the other hand, objects with {\em many} degrees of freedom will almost always be classical, their quantum potential being exponentially suppressed, and all the ``worlds'' in the ensemble evolving as separate, essentially noninteracting entities -- our tennis ball being one of them. In other words, an ensemble of such objects behave just like a mixed quantum state. Recall, that in a standard density matrix approach a mixed state is produced by a process of decoherence, which exponentially dampens all non-diagonal elements in the density matrix, resulting in a set of classical alternatives -- a.k.a. a set of classical independent ``worlds'' in the MIW framework.

\subsection{An Overture for a MIU}\label{Sec:MMIIUU}

By this point an attentive reader might have already surmised the kind of corollary which immediately follows from our discussion: while the quantum effects are essentially unobservable at the macroscopic scale -- even for such tiny objects as bacteria and viruses\footnote{The smallest known bacteria, {\em Pelagibacter ubique}, has an average cell diameter of $1.2-2 \times 10^{-7}$ meters, whereas the smallest known virus, {\em Porcine circovirus} (a single-stranded DNA virus) has a capsid with approximate diameter of $1.7\times 10^{-8}$ meters.}, and even more so for humans, planets, stars and galaxies -- they might arise with a vengeance once we zoom out onto a very large, cosmological scale! The reason for this is simple: after all, the observable Friedman universe possesses the same amount of degrees of freedom as a single material particle in a gravity field. Even though at preset it contains around  $10^{80}$ elementary particles, unevenly spread out and clusterized into various galaxies, once we reach the scales larger than 250/h Mpc\footnote{Due to uncertainty in the exact value of Hubble parameter $H$, for sufficiently large distances a dimensionless Hubble constant $h$ is commonly used, $h\in (0.5,0.75)$.}, the universe smoothes out to an incredible degree of homogeneity  and isotropy -- a long-lasting endowment of cosmological inflation in the early universe. This observation, taken together with MIW formalism, naturally brings us to the following question: is it possible for a {\em contemporary} Friedman universe to experience quantum effects? And if yes, is there a way to detect it?

A partial answer to this question has been proposed in \cite{YY}. In that article we have pointed out the known fact that the Friedman cosmology can be derived as a simple generalization of a Newtonian dynamics in a centrally symmetric gravity field, which makes the Friedman universe uniquely susceptible for quantization via the MIW approach. In fact, it appears to be the most {\em natural} target for MIW formalism; after all, for an ensemble of $\mathcal{D}$ interacting particles (a.k.a. ``worlds''), the complete convergence between the non-relativistic quantum mechanics and MIW occurs only in the continuous limit of $\mathcal{D} \to \infty$. But it is not longer the case for the cosmology! For example, in the accelerating dS model (in the framework of $\Lambda {\rm CDM}$-model) we are enclosed by a cosmological event horizon. Using the holographic principle \cite{Hooft1993}, \cite{Sussk1994}, \cite{Bek1972} we must therefore allot a large -- very large! -- but nevertheless {\em finite} number of possible quantum states for the observable universe: something on the order of $\exp(10^{122})$. But this also limits the amount of discernible classical ``worlds'', which ends up being finite on a purely fundamental way, and not just as a mere approximation. Similar reasoning can be conducted even in the absence of cosmological acceleration, since the distances exceeding $c/H$ (where $H$ is a Hubble parameter) are unobservable as a matter of principle (also it must be noted, that in non-accelerating Friedman cosmology $H$ decreases with time, thereby accruing the total number of possible quantum states). To sum up, the MIW approach in cosmology is not only warranted, but is in general more fundamental than in standard quantum mechanics.

The current article follows the ideas proposed in  \cite{YY}, but can also be read as an independent work. In the course of the article we are planning to show that the formalism of Many Interacting Worlds -- which in our case should be more properly called the formalism of Many Interacting Universes (MIU) -- allows one to formulate and subsequently prove the following theorem: the formalism of MIU allows to completely dispense with the cosmological singularities of Big Bang, Big Crunch and even Big Rip types. Interestingly, the key role in the model is played by a special type of Friedman universes, first introduced in the literature by professor John Barrow.

On September 26, 2020 the world of cosmology was shocked by the news of death of a prominent theoretical physicist, brilliant mathematician and indomitable popularizer of science, professor John David Barrow. This was a terrible and irretrievable loss, for professor Barrow had always been a source of incredibly novel, visionary ideas, and had possessed a wonderfully accurate physical intuition which allowed him to immediately discern and appreciate the most interesting ideas of his colleagues. One of the authors of this article (Artyom) has had a pleasure of many years of correspondence with him, and can attest that John was never shy to offer his friendly support and good advices. Many of our articles would have never seen the light of day if not for his encouragement. The loss of John Barrow is a great tragedy not just for his family but for all of us...

In the last year of his life Professor Barrow has published as many as 18 (!) articles in ArXiv, all of them written in a lucid, academic style, so characteristic of his deeply perceptive mind. Many of those articles contains entirely new ideas and concepts that are sure to bear fruits in the future. We will concentrate on just one of those articles, \cite{Zero-Univ}. In it, Professor Barrow has introduced unusual (ill-posed) cosmological models where both the scale factor and its first derivative are always equal to zero:
\begin{equation}
a(t)={\dot a}(t)=0,\qquad  \forall\,\, {t}
\label{ZU}
\end{equation}
where the overdot denotes the derivative w.r.t. time. The existence of unusual solutions like these is predicated upon the ill-posedness of initial conditions due to a lack of a local Lipschitz condition. To illustrate this, consider a scale factor describing a power-law inflation: $a(t)=t^n$, with $n>1$. Such a solution of Einstein equations in a homogeneous isotropic universe satisfy the very simple initial conditions:
\begin{equation}
a(0)={\dot a}(0)=0.
\label{IC-ZU}
\end{equation}
However, the unusual solution (\ref{ZU}) also happens to satisfy them. Thus, we are facing a problem: we cannot dispense with the power-law inflation solutions, but if we are to stick with them, we must be able to explain why a violation of the Lipshitz condition might be perfectly acceptable (in the case of a power-law inflation, the Lipshitz condition is violated by an unlimited growth of the first derivative of scale factor). Granted, we might dismiss the solutions (\ref{ZU}) by painting them as essentially unphysical, but it would hardly solve the problem. After all, are we really prepared to discard something that is a crucial ingredient of the cosmological models of Vilenkin \cite{Vil1982}, where the universes are literally created from ``nothing'' -- i.e. sprawl from the ``zero universe''? These considerations lead us to the question formulated by Barrow at the very end of his article \cite{Zero-Univ}: ``{\em Creation out of nothing may create something that does not turn into a universe as we understand it. What does this mean?...}''.

In this article we are going to take a look at the Barrow's zero universes from yet another angle. We are planning to demonstrate that these universes, after being accounted for in the framework of a quantum theory formulated in the MIU approach, and under some physically sensible assumptions, lead us to a remarkable conclusion: the zero universes stabilize the multiverse by preventing the cosmological singularities from ever forming! In other words, it is possible that Professor Barrow had discovered a crucial cosmological element that might help theoreticians to achieve the old and seemingly unreachable goal: to prove that the cosmological singularities -- not only the standard big bang/big crunch types, but the big rip as well -- are merely the relics of the classical theory and as such will be eliminated in a proper quantum field theory.

The article is structured as follows: in Sec. \ref{Sec:EQ} we will show that (\ref{ZU}) can indeed be explicitly derived as an exact solution for the Friedman equations, and is in fact a special case of a famous solution known as the Einstein quintessence \cite{EQ}. In Sec. \ref{Sec:MIU} we provide the short explanation of MIU approach and formulate a set of conditions that will be necessary to ensure these models are free of singularities. In Sec. \ref{Sec:no-singularity} we put these conditions to use by proving a Theorem \ref{Theorem1} about the non-existence of singularities. This statement of the theorem is illustrated by a few simple completely integrable examples in Sec. \ref{Sec:exact-solutions}, where we separately consider the ensembles of two and three universes -- the latter one studied using the Hamilton-Jacobi formalism.
Next section, Sec. \ref{Sec:Phantom}, is dedicated to the problem of Big Rip singularities and how they can be avoided thanks to the MIU-style quantization. Finally, Sec. \ref{Sec:discussion} serves as a conclusion of the article. In it we first briefly discuss the extension of the proposed formalism on the inhomogeneous and anisotropic models, and then propose a hypothesis that by adding the zero universe to the MIU approach one not only gets rid of cosmological singularities, but also opens a possible pathway to solve yet another two fundamental problems in the cosmology of an early universe: the decoherence problem and the problem of a low-entropy initial state which is necessary for the cosmological inflation to commence. Also, for the interest of completeness we have included the Appendix \ref{sec:Appendix}, in which we elucidate how the search for a physically meaningful interpretation of the mathematical laws of quantum mechanics has led the physicists first to the formulation of de Broglie-Bohm interpretation, then to its upgraded version, the Many Interacting World (MIW) approach by Hall, Deckert and Wiseman, which in turn has led to the cosmological model of Many Interacting Universes, a keystone of the current article.

\section{The Barrow universe's essence is nothing but Einstein quintessence}\label{Sec:EQ}

In this section we will show that the Barrow's zero universes (\ref{ZU}) cannot be dismissed as an ill-posed mathematical artifact and should be -- or, at the very least, can be -- treated as possible, physically permissible universes. This is an important point even by itself, especially for the quantization problems in the framework of functional integrals, in which the probability amplitudes are determined via the integration along {\em all physically consistent metrics}, and must therefore necessarily include the metrics similar to (\ref{ZU}). We will not pursue this line of reasoning in this article, however; instead we are going to propose a different approach to the same quantization problem (the MIU method), which not only allows to naturally account for zero universes (\ref{ZU}) but also ensures that (at least in the MIU framework) the quantum gravity within the Friedman metrics is in fact completely emancipated from the threats of Big Bang, Big Crunch and even Big Rip singularities! But before we get there, we have to establish some important facts about the solutions akin to \eqref{ZU}.

Let us begin by generalizing (\ref{ZU}) as follows:
\begin{equation}
a(t)=\frac{ d^n a(t)}{dt^n}=0,\qquad  \forall\,\, {t},\,\, \forall\,\, {n\in Z}.
\label{ZU-gen}
\end{equation}
This is, in fact, a natural extension of \eqref{ZU}: after all, if \eqref{ZU-gen} is violated, and there exists a moment of time $t_1$ when at least one of the higher derivatives $d^n a/dt^n$ pops out of a zero background, then there would necessarily follow a moment $t_2>t_1$, arbitrarily close to $t_1$, when the derivative $d^{n-1} a/dt^{n-1}$ is to become non-zero. The same fate would then befall the $(n-2)$-nd derivative, followed by the $(n-3)$-rd derivative, and so on. The resulting domino effect will necessarily culminate in a violation of the condition (\ref{ZU})\footnote{In fact, if we assume that $a^{(n)}(t) = \epsilon(t)$ while all lower order derivatives of $a(t)$ and the scale factor itself are equal to zero at some $t=t_1$, then we can use the Cauchy formula for repeated integration to get the formulas:
\beqn
a(t) = \frac{1}{(n-1)!}\int\limits_{t_1}^{t} (t-\xi)^{n-1} \cdot \epsilon(\xi) d \xi, \qquad \dot a(t) = \frac{1}{(n-2)!}\int\limits_{t_1}^{t} (t-\xi)^{n-2} \cdot \epsilon(\xi) d \xi,
\enqn
that obviously violate \eqref{ZU-gen} if $\epsilon(t) \neq 0$ for $\forall t$.}, thus motivating and explaining our interest in the condition (\ref{ZU-gen}). Hence, the main goal for this section would be to demonstrate that the universes that satisfy this condition naturally arise as special solutions of the Friedmann-Lema\^itre-Robertson-Walker (FLRW) equations.

For the sake of simplicity we will restrict ourselves to the universes filled by the fields of matter with the barotropic equation of state $w=p/\rho c^2={\rm const}$, where $\rho$ is the density, $p$ is the pressure and $c$ denotes the speed of light. These variables are related to the scale factor $a$ via the O.D.E.
\beq \label{rho_dot}
\dot\rho = -3 \left(\rho + \frac{p}{c^2}\right) \frac{\dot a}{a},
\enq
which for the barotropic equation of state has the general solution
\begin{equation}
\alpha\rho=a^{-3(w+1)},\qquad \alpha={\rm const}.
\label{rho-al}
\end{equation}
Let us plug \eqref{rho-al} into the FLRW equations:
\begin{equation}
\begin{split}
\dot a^2 &=\frac{8\pi G}{3}\rho a^2 - k c^2,\\
{\ddot a} &=-\frac{4\pi G}{3}\left(\rho + \frac{3p}{c^2}\right)a,
\label{FLRW-start}
\end{split}
\end{equation}
where $k=0,\,\pm 1$ is the curvature constant. This results in:
\begin{equation}
\begin{split}
\dot a^2 &=\frac{8\pi G}{3 \alpha} a^{-(1+3w)} - k c^2,\\
\ddot a &=-\frac{4\pi G}{3\alpha}\left(1+3w\right)a^{-(3w+2)}.
\label{FLRW_0}
\end{split}
\end{equation}
Let's take a look at the solutions of \eqref{FLRW_0} under two necessary requirements: that they describe a collapsing universe, and that $a\to 0$ and ${\dot a}\to 0$ as $t\to t_s$  in accordance with (\ref{ZU}). If $k\neq 0$ (the case of a non-flat universe), these requirements can only be satisfied if
\begin{itemize}
\item the universe is closed ($k=+1$) and
\item $w= -1/3$.
\end{itemize}
Such a solution has been named the {\em Einstein quintessence} (EQ) \cite{EQ}, and it is known to contain a Tipler's Omega point \cite{Omega}. Also, the constant $\alpha$ must be restricted to the interval:
\begin{equation}
0<\alpha\le \frac{8\pi G}{3 c^2}.
\label{ner-al}
\end{equation}
Since we are talking about a collapsing universe, let us set ${\dot a}<0$. The general solution of (\ref{FLRW_0}) for a closed unverse with $w=-1/3$ has the form:
\begin{equation}
\begin{split}
a(t) & = \left(t_s-t\right)\cdot \sqrt{\frac{8\pi G}{3\alpha}-c^2},\\
\frac{da(t)}{dt} &=-\sqrt{\frac{8\pi G}{3\alpha}-c^2},\qquad \frac{d^n a(t)}{dt^n}=0, \,\,\,\,n>1,
\label{EQsol}
\end{split}
\end{equation}
where the constant of integration $t_s$ determines the time of arrival of the final singularity. And now we can make the key observation: by setting $\alpha$ equal to the upper limit of inequality (\ref{ner-al}), the solution (\ref{EQsol}) turns into (\ref{ZU-gen}), exactly as we required. To put it in other words, Barrow's zero universe is simply a special case of (\ref{EQsol}), and has a structure of the Tipler's  Omega point~\footnote{Note, that the condition (\ref{ner-al}) ensures there to be no event horizons: the universe with the Tipler's Omega point does not experience the informational paradox! \cite{Omega} Furthermore, the solution can be generalized for the broad class of closed, spherically symmetric classes of space-time manifolds.}.

If we introduce a final constant $0<a_s<\infty$
\begin{equation}
t_s=a_s\left(\frac{8\pi G}{3\alpha}-c^2\right)^{-1/2},
\label{tsas}
\end{equation}
and then take a limit $\alpha\to 8\pi G c^{-2}/3$, we'll have $t_s\to \infty$ and (\ref{EQsol}) will turn into a stationary Einstein solution
\begin{equation}
a(t)=a_s,\qquad \frac{ d^n a(t)}{dt^n}=0,\qquad  \forall\,\, {t},\,\, \forall\,\, {n\in Z},
\label{EQ-gen}
\end{equation}
which explains the term ``Einstein quintessence'' chosen for this model (see \cite{EQ}). Hence, a zero universe (\ref{ZU-gen}) is indeed a special case of EQ (\ref{EQ-gen}) with $a_s=0$.

 But is it the {\em only} model where it might naturally arise? Can we not gain a zero universe (\ref{ZU-gen}) by some other means besides the reduction from EQ? To answer this question let us first reiterate that the open universe case (when $k=-1$) must be discarded right away, owing to the requirement $\rho \ge 0$. This leaves us the flat universes with $k=0$. For them both ${\dot a}\to 0$ and $a\to 0$ only when $-1<w<-1/3$ (the phantom case $w<-1$ deserves a separate discussion, and will be granted one in Sec. \ref{Sec:Phantom}). Integrating (\ref{FLRW_0}) leads to:
\begin{equation}
a(t)=\left[\left(w+1\right)\sqrt{\frac{6\pi G}{\alpha}}\left(t_s-t\right)\right]^{\frac{2}{3(w+1)}}.
\label{NEQsol}
\end{equation}
Here we are facing an interesting problem. Unlike (\ref{EQsol}), the solution (\ref{NEQsol}) is irreducible to $a(t) \equiv 0$ except when $\alpha\to\infty$. In order to see this, first consider the case when the power in the r.h.s. of (\ref{NEQsol}) is not an integer: $n=2/(3(w+1)) = N +\epsilon$, where $N\in \N$, $\epsilon \in (0,1)$. Then {\em every} higher order derivative of (\ref{NEQsol}) starting with $N+1$'st, will necessarily diverge at $t\to t_s$ for $\alpha<\infty$.
On the other hand, if $n\in Z$ and the parameter of state $w$ has the form:
\begin{equation}
w=-1+\frac{2}{3n},
\end{equation}
then the $n$-the derivative of the scale factor will be a {\em non-zero} constant for all finite $\alpha$, in complete violation of \eqref{ZU-gen}. The only way to force this universe to comply with \eqref{ZU-gen} is by choosing $\alpha\to \infty$, which, according to \eqref{rho-al}, means a completely {\em empty} universe. Which is  physically possible except that in a standard (non-stationary) Friedman cosmology an empty universe must be open, and not flat. In other words, we cannot claim the solution \eqref{NEQsol} to be fully compliant with (\ref{ZU-gen}), as we did for the case (\ref{EQsol}). Fortunately, the latter is sufficient to prove our point: that the solution (\ref{ZU-gen}) is not a mere mathematical artifact and is physically permissible, just like the stationary Einstein solution (\ref{EQ-gen}), whose special case it happens to be. And while this does not lead to any interesting consequences in the classical gravity theory, within the quantum MIU formalism it has some very serious ramifications and becomes a necessary (if unlikely) ingredient in a special singularity sequestering stew, which we will soup up in the next section.


\section{Barrow Universe and The Multiverse of Sanity} \label{Sec:MIU}
The MIU model was originally suggested in \cite{MIW} where it was called the Many Interacting Worlds model and was designed as an interpretation of the quantum mechanics which united the best qualities of the de Broglie-Bohm pilot wave interpretation \cite{Bohm} and the Many-Worlds Interpretation (MWI) \cite{MWI} (for the history and mathematics behind both of them See Appendix \ref{sec:Appendix}). The ramification of this model for cosmology were explored in \cite{YY}. We won't get into details of derivation of that model here -- we provide that in Appendix \ref{sec:Appendix} for the interested reader -- but instead concentrate on the general gist of the MIU approach.

The main idea behind the MIU model is that instead of a single universe there exist an entire ensemble of them, just like in the MWI interpretation. However, unlike the latter, the universes in MIU are interacting with each other. This interaction is governed by a following quantum potential:
\begin{equation}
U(a_1,...,a_{_N})=\sum_{n=1}^N\left(\frac{1}{a_{n+1}-a_n}-\frac{1}{a_n-a_{n-1}}\right)^2,
\label{U}
\end{equation}
where $a_n$ denotes the scale factor of the $n$-th universe and the ensemble consists of exactly $N$ mutually interacting universes. For consistency, we also have the boundary condition:
\begin{equation}
a_0=a_{_{N+1}}=\infty.
\label{bc}
\end{equation}

The interaction between the ``neighbouring'' universes (i.e. the universes with sufficiently close values of scale factors) leads to a dynamics that is controlled by the $N$ Lagrange-Euler equations \cite{YY}:
\begin{equation}
\begin{split}
{\ddot a}_n=&-\frac{4\pi G}{3}\left(\rho_n+\frac{3p_n}{c^2}\right)a_n-\\
&-\left(L_{_{PL}}c\right)^2\frac{\partial}{\partial a_n}U(a_1,...,a_{_N}),
\end{split}
\label{LE}
\end{equation}
and is additionally constrained by the following equation:
\begin{equation}
\begin{split}
\sum_{n=1}^N & \left(\frac{1}{2} {\dot a}_n^2-\frac{4\pi G}{3}\rho_na_n^2+\frac{c^2}{2}k_n\right)+ \\
&+\left(L_{_{PL}}c\right)^2U(a_1,...,a_{_N})=0,
\end{split}
\label{con}
\end{equation}
where $\rho_n=\rho_n(a_n)$ is the density of matter in the $n$-the universe (and it satisfies the usual continuity equation \cite{YY}), $p_n=p_n(a_n)$ denotes the pressure, constants $k_n=-1,\,0,\,1$ are the Gaussian curvatures, and for simplicity we have introduced an additional Plank multiplier:  $L_{_{PL}}=(8\pi \hbar G/3c^3)^{1/2}$. Thus, in order to understand the cosmological dynamics in the resulting multiverse we have to integrate the equations (\ref{LE}) with the constraint (\ref{con}) for the $N$ given equations of state and initial conditions.

However, a close consideration of the problem reveals one additional complication. As we have discussed in \cite{YY}, from a physical point of view it is necessary for very small universes to experience strong influence of quantum effects. In other words, when $a_n\to 0$, the quantum phenomena must necessarily manifest themselves, if not dominate the dynamics altogether. But in the framework of MIU all quantum effects have but a single source: {\bf the quantum potential} $U(a_1,..)$. The only  possible conclusion would be that for $\forall\,n$  as $a_n\to 0$ the quantum potential ought to diverge. And here we have a problem, because generally speaking, it does not -- see (\ref{U}). One possible workaround that helped to overcame this problem was to introduce a ``master-factor'' $a(t)$, that all the scale factors in the multiverse are proportional to, i.e. assume that for $\forall\,t$: $a_n(t)=\mu_n a(t)$, where $\mu_n$ denotes a positive dimensionless constant associated with an $n$-th universe \cite{YY}. Admittedly, this is essentially an ad hoc assumption, introduced to solve the aforementioned problem with the small universes. Fascinatingly, this very problem vanishes completely when we add to the ensemble of interacting universes a single Barrow's zero-universe! Moreover, as we shall see below, the presence of the corresponding term in (\ref{con}) under some very simple conditions produces such a fast growth of $U$ as $a_n\to 0$ that the resulting quantum repulsion effectively prevents the cosmological singularity from occurring altogether. Before we move on to proving this extraordinary statement, we shall discuss the conditions that are necessary to make it true.

\begin{condition}\label{condition1}. At some time $t=t_0$ (for simplicity we will henceforth assume that $t_0=0$) the values of all scale factors of the universes are ordered according to their index numbers:
\begin{equation}
a_n(0)>a_{n-1}(0).
\label{C1}
\end{equation}
\end{condition}

\begin{condition}\label{condition2}. The first (and therefore the smallest) universe in the ordered list of scale factors $\{a_n\}$ is the Barrow's zero-universe satisfying (\ref{ZU}):
\begin{equation}
a_1(t)={\dot a}_1(t)=0,\qquad {\forall\,\,t}.
\label{C2}
\end{equation}
\end{condition}

\begin{condition}\label{condition3}. The second universe in the list (\ref{C1}) with a scale factor $a_2$ is an entirely quantum universe containing no fields of matter:
\begin{equation}
\rho_2=p_2=0.
\label{C3}
\end{equation}
\end{condition}

\begin{condition}\label{condition4}. The fields of matter in the rest of the universes with $n>2$ satisfy the standard energy conditions employed in the theorems about the singularities \cite{How}. In particular, we assume that  $\rho_n$ has but one special point that occurs only when $a_n\to 0$. ~\footnote{The only exception are the Big Rip singularities which will be dealt with in Sec. \ref{Sec:Phantom}.}
\end{condition}

Let us recall that the densities, pressures and scale factors in every universe satisfy the continuity condition \cite{YY}:
\begin{equation}
{\dot\rho}_n=- 3\frac{\dot a_n}{a_n}\left(\rho_n+\frac{p_n}{c^2}\right).
\label{ner}
\end{equation}
By using (\ref{LE}) together with (\ref{ner}) it is easy to see that in the classical limit $\hbar\to 0$ ($L_{_{PL}}\to 0$) the equations (\ref{LE}), (\ref{con}) are reduced to a system of very familiar cosmological equations:

\begin{equation}
\begin{split}
\left(\frac{{\dot a}_n}{a_n}\right)^2=\frac{8\pi G}{3}\rho_n-\frac{k_n c^2}{a_n^2},\\
{\ddot a}_n=-\frac{4\pi G}{3}\left(\rho_n+\frac{3p_n}{c^2}\right)a_n,
\label{FLRW}
\end{split}
\end{equation}
for $n>2$ and
\begin{equation}
{\ddot a}_2=0,
\label{clas}
\end{equation}
which describes a system of $N$ classical non-interacting Friedmann-Lema\^itre-Robertson-Walker (FLRW) universes. This little exercise effectively demonstrates that the MIU formalism is essentially just a different method of quantization, whereas the process of taking the limit $\hbar\to 0$ plays the role of decoherence. Indeed, what is a decoherence if not a process during which every non-diagonal element in the density matrix vanishes exponentially fast? And this means that the decoherence eradicates the interference terms and thus converts the {\em superpositions} into the {\em mixtures}, rapidly producing an ensemble of mutually non-interacting ``universes'' -- exactly the same result as the one naturally following from the MIU approach.

However, an attentive reader might ask the following question: ``So, the modified equation can indeed be reduced to the classical FLRW equations. But does it necessarily mean that the {\em solutions} of MIU system are smoothly reducible to the usual  FLRW solutions? In other  words, if we take a limit $\hbar\to 0$ -- would the solution be {\bf perturbative} with respect to $\hbar$ (in our case to $L_{_{PL}}$)?'' This question becomes even more pressing if we recall that most of the solutions constructed in  \cite{YY} do not satisfy that property due to the fact that they were produced using the ``master-factor''  condition -- the condition that ill reconcilable with the system (\ref{LE}). But fortunately, the answer to this question is ``yes''! As we shall see in Sec. \ref{Sec:exact-solutions}, the perturbativity is indeed present in our solutions and so is the possibility of decoherence. It is as well, since the absence of the classical limit in the model makes it rather suspicious for our tastes -- although we fully admit that there might be different opinions regarding this matter, so we will abstain from discussing it in this article.

\section{The Conspicuous Absence of Singularities}\label{Sec:no-singularity}

Before we begin formulating the main Theorem that rules out singularities in the proposed model, it will be a good idea to get a bit more acquainted with the quantum potential (\ref{U}). It depends on $N$ scale factors $a_n$, whose appearances follow a very simple pattern: every $a_n$ with indices $2<n<N-1$ is present in just three terms:
\begin{equation}
\begin{split}
U(a_1,...,{\bf a_n},...a_{_N})=...&+\left(\frac{1}{{\bf a_n}-a_{n-1}}-\frac{1}{a_{n-1}-a_{n-2}}\right)^2\\
&+\left(\frac{1}{a_{n+1}-{\bf a_n}}-\frac{1}{{\bf a_n}-a_{n-1}}\right)^2+\\
&+\left(\frac{1}{a_{n+2}-a_{n+1}}-\frac{1}{a_{n+1}-{\bf a_n}}\right)^2+....
\end{split}
\label{an}
\end{equation}
The scale factors $a_2$ and $a_{_N}$ are a bit different in this regard.  In order to see how they enter the potential (\ref{U}) we will need to take (\ref{bc}) and (\ref{C2}) into account, producing the following relationship:
\begin{equation}
\begin{split}
U(a_1,{\bf a_2},...,{\bf a_{_N}})=&\frac{1}{{\bf a_2}^2}+\left(\frac{1}{a_3-{\bf a_2}}-\frac{1}{{\bf a_2}}\right)^2+\left(\frac{1}{a_4-a_3}-\frac{1}{a_3-{\bf a_2}}\right)^2+...\\
\\
&...+\left(\frac{1}{{\bf a_{_N}}-a_{_{N-1}}}-\frac{1}{a_{_{N-1}}-a_{_{N-2}}}\right)^2+\frac{1}{({\bf a_{_N}}-a_{_{N-1}})^2}.
\end{split}
\label{UU}
\end{equation}

Now that we are familiar with the structure of \eqref{U}, let us not waste time and take a look at the main Theorem:
\begin{theorem} \label{Theorem1}
If the conditions \ref{condition1}-\ref{condition4} are satisfied, the solutions of the system (\ref{LE}) and (\ref{con}) are strictly positive for all $n > 1$.
\end{theorem}

{\bf The Proof}. First of all, it is easy to see that the scale factor $a_2$ of the purely quantum universe (see condition \ref{condition3}) will always be non-zero as long as for $\forall\,n>2$ a strict inequality
$a_n>0$ holds. This follows directly from (\ref{UU}) and (\ref{con}): if $a_2\to 0$, then $U(0,a_2,..,a_{_N})\to +\infty$; the only way to compensate for it would be by a proper growth of $\rho_n$, which is forbidden by (\ref{C3}). Hence, a singularity in the second universe would imply an unlimited growth of the l.h.s. of constraint (\ref{con}), duly violating this very constraint in the process.

This conclusion can be extended for the remainder of the multiverse: for any $n\neq 1$ the scale factor $a_n$ cannot turn to zero as long as $a_i>0$ with $1< i < n$. In order to prove this, assume that at some $t'$ a universe with $n>2$ experiences a singularity $a_n(t')=0$, while every other unverse does not. Then, according to (\ref{C1}) there exists such moment of time $0<t''<t'$ when $a_{n-1}(t'')=a_n(t'')>0$. However, the particular form of (\ref{U}) ensures that at $t''$ the quantum potential becomes divergent:
$$
\displaystyle{
U(0,a_2(t''),...,a_{n-1}(t'')=a_n(t'')>0,...,a_{_{N}}(t''))=+\infty}
$$
once again violating the constraint (\ref{con}), since by the condition \ref{condition4} the rest of the terms in (\ref{con}) remain finite.

Therefore, if $a_n$ turns to zero, the same fate shall befall all $a_k$ with $k<n$, and it should happen simultaneously. Indeed, if we assume the contrary i.e. that for all $a_k$ there exists such times $t_k$ that $a_k(t_k)=0$, then according to our reasoning those moments have to be ordered as such: $t_2<t_3<...<t_n$. Hence, at $t_2$ we should have both $a_2(t_2)=0$ and $a_k(t_2)\ne 0$ for all $k>2$, which once again violates the constraint (\ref{con}).

So, the scale factor $a_n$ for any given $n>1$ might turns to zero at some time $t'$ only simultaneously with every other $a_k$ with $k<n$:
\begin{equation}
a_2(t')=a_3(t')=...=a_{n-1}(t')=a_n(t')=0,
\label{master}
\end{equation}
where
\begin{equation}
a_k(t)<a_{k+1}(t),\qquad \forall\,\, {2\le k\le n}, \qquad t \neq t'.
\label{mas1}
\end{equation}
For the sake of simplicity let $t'=0$ \footnote{This can be done without any loss of generality thanks to the invariance of \eqref{LE} and \eqref{con} to the time translations and time inversions}. From the point of view of functions $\{a_n\}_{n=2}^N$ there are two ways to satisfy (\ref{master}) and (\ref{mas1}) at $t=0$: (i) as $t\to 0$ we should have $a_k/a_{k+1}\to 0$ for $k=2,..,n-1$ or (ii) $a_k/a_{k+1}\to {\rm const}$. In the former case in the close vicinity of $t=0$ the functions $a_n$ behave as a set of power functions:
\begin{equation}
a_k(t) \propto \mu_k t^{m_k}, \quad 0<m_k<1,\quad m_{k+1}<m_k,
\label{mas2}
\end{equation}
where  $\mu_k={\rm const}$ and the exponents $m_k \in (0,1)$ in order to satisfy the Conditions \ref{condition1} and \ref{condition4}. Case (ii) is essentially a variation of a ``master-factor hypothesis'' from \cite{YY}, localized in a sufficiently small vicinity of $t=0$, and it implies the existence of such function $a(t)$ and such dimensionless constants $\mu_k$, $k=1,...,\,n$, that
\begin{equation}
\lim_{t\to 0} a_k(t)=\mu_k a(t),\qquad \mu_{k+1}>\mu_k.
\label{master1}
\end{equation}
Now let's consider both of these cases, beginning with the first. In case (i) as $t\to 0$
$$
\frac{1}{a_{n+1}-a_n}-\frac{1}{a_n-a_{n-1}}\to \frac{1}{a_{n+1}}-\frac{1}{a_n}\to -\frac{1}{a_n},
$$
so at a sufficiently close vicinity of $t=0$ the dominant role in the potential (\ref{U}) is played by the scale factor $a_2$:
\begin{equation}
U \approx U(a_2)=\frac{2}{a_2^2}.
\label{U0}
\end{equation}
On the other hand, $a_2(t)$ satisfies the system (\ref{LE}) with $\rho_2=p_2=0$, in particular, the equation:
\begin{equation}
{\ddot a}_2=-\left(L_{_{PL}}c\right)^2\frac{dU(a_2)}{d a_2}.
\label{LE2}
\end{equation}
This differential equation allows for a reduction of order and is therefore completely integrable. Its first integral and the general solution are of the form:
\begin{equation}
\begin{split}
\frac{{\dot a_2}^2}{2} &+\frac{2 L^2_{_{PL}}c^2}{a_2^2}=\epsilon^2>0, \\
a_2(t) &=\sqrt{2\epsilon^2\left(t-t_0\right)^2+\frac{2 L^2_{_{PL}}c^2}{\epsilon^2}},
\end{split}
\label{aa2}
\end{equation}
where $\epsilon^2$ and $t_0$ are two real-valued constants of integration. \eqref{aa2} unequivocally proves that $a_2(t)$ cannot turn to zero for {\em any} value of $t$, directly contradicting our initial assumption and thus proving our theorem for the Case (i)~\footnote{It is also possible to prove our theorem by utilizing (\ref{mas2}) directly; the direct computation $0<m_2<1$ would yield the analogue of (\ref{LE2}) where the l.h.s. is always negative whereas the r.h.s. is strictly positive, thus producing the thought after discrepancy}.

The Case (ii) can be tackled using the same strategy, i.e. by deriving the equation for the local ``master-factor'' $a(t)$ from (\ref{master1}), integrating it, and directly verifying that the solution (\ref{aa2}) is once again a strictly positive function. We will omit the calculations since they are again rather straightforward albeit somewhat more cumbersome than in Case (ii).

\section{The Taming of the Few: Some Exact Solutions}\label{Sec:exact-solutions}

While, generally speaking, the equations (\ref{LE}) and (\ref{con}) are non-integrable, we can nevertheless wrangle out at least some exact solutions for the simpler cases of $N=2$ and $N=3$ interacting universes with a select choice of matter densities. Our goal for this Section will be to illuminate the behaviour of scale factors and to compare their behaviour with the predication of Theorem \ref{Theorem1}. As an added bonus, we will also try to set the record straight about the perturbativity of solutions and discuss the existence of a classical limit $L_{_{PL}}\to 0$.

\subsection{A tale of two universes: $N=2$.}\label{ssec:N2}

For the first example let us consider a limited multiverse consisting of just two universes: the zero universe with $a_1=0$ and a universe, whose scale factor for brevity will be denoted by $x$: $a_2(t)=x(t)$. Our problem is then reduced to integrating one equation (\ref{con}):
\begin{equation}
\frac{1}{2}{\dot x}^2-\frac{4\pi G}{3}\rho(x)x^2+\frac{2c^2L_{_{PL}}^2}{x^2}+\frac{c^2 k}{2}=0,
\label{con2}
\end{equation}
where $k=k_2$. Because this is a rare situation where we are able to derive and savour all the possible solutions, we will temporarily abandon the previously stated Condition \ref{condition3} and consider a whole gamut of possible densities $\rho(x)$. Naturally, all of them (except when $\rho(x) \equiv 0$) would blatantly violate the condition of the Theorem \ref{Theorem1}, opening a possibility of emergent singularity whenever the parameter of state $w$ satisfies the inequality
\begin{equation}
w=\frac{p}{\rho c^2}\ge \frac{1}{3},
\label{sing}
\end{equation}
since this condition at $x\to 0$ ensures the domination of the second term in (\ref{con2}) over the quantum repulsion when (\ref{sing}) is a strict inequality (when $w=1/3$ it settles for a near-domination).

But we have to start somewhere. Let us begin by considering the simplest case possible.

\begin{itemize}

\item The purely quantum universe: $\rho(x)=0$. In this case (\ref{con2}) has a real-valued solution only for the open model, $k=-1$. The solution itself has the form:
\begin{equation}
x(t)=\sqrt{c^2\left(t-t_0\right)^2+4 L_{_{PL}}^2},
\label{solution-1}
\end{equation}
where $t_0 \in \R$. This is exactly the same solutions that has been previously constructed in \cite{YY} (see also (\ref{aa2}) with the aid of the ``master-factor'' hypothesis. Apparently, this hypothesis is no longer needed as long as we include the Barrow's zero universe!

\item The universe filled by the baryon matter (a.k.a. the ``dust'') with $\rho(x)=D^2/x^3$. The curvature $k$ is not fixed, leading to three alternatives:

{\bf (i)  A closed model, $k=+1$}. The exact solution has the form
\begin{equation}
\begin{array}{l}
\displaystyle{
x(\eta)=\mu+\sqrt{\mu^2-4 L_{_{PL}}^2}\sin\eta,}\\
\\
\displaystyle{
t(\eta)=t_0+\frac{1}{c}\left(\mu\eta -\sqrt{\mu^2-L_{_{PL}}^2}\cos\eta\right),}
\end{array}
\label{solution-2pl}
\end{equation}
with  $\mu=4\pi G D^2 /(3c^2)$. If we take a limit at $\hbar=0$, this solution turns into a familiar Friedman solution for the closed universe filled with dust. In general, though, our solution contains no final singularity; instead, it describes an eternally oscillating universe with periodic rebounds occurring at $\eta=-\pi/2+2\pi m$ followed by the periods of expansion lasting until $\eta=\pi/2+2\pi m$. The maximal expansion of the universe is $x_{max}=\mu + \sqrt{\mu^2- L_{_{PL}}^2}$ and the minimal possible value for the scale factor is $x_{min}=\mu - \sqrt{\mu^2- L_{_{PL}}^2}$. It is also interesting to note that the period of oscillations does not depend on $L_{_{PL}}^2$ and is equal to $2\mu\pi/c$.

{\bf (ii) An open model, $k=-1$}. In this case the solution has the following parametric form:
\begin{equation}
\begin{array}{l}
\displaystyle{
x(\eta)=\sqrt{\mu^2+4 L_{_{PL}}^2}\cosh\eta-\mu,}\\
\\
\displaystyle{
t(\eta)=t_0+\frac{1}{c}\left(\sqrt{\mu^2+L_{_{PL}}^2}\sinh\eta-\mu\eta \right).}
\end{array}
\label{solution-2mn}
\end{equation}
At $L_{_{PL}}=0$ (\ref{solution-2mn}) we again get a well-known Friedman solution and it is singular at $t=t_0$ ($\eta=0$); it no longer is if $L_{_{PL}}>0$.

{\bf (iii) A flat model, $k=0$}. After integrating (\ref{con2}) we attain
\begin{equation}
\begin{split}
\pm t = t_0+&\sqrt{\frac{2}{\mu c^2}}\left(\frac{1}{3}\left(x-\frac{2L_{_{PL}}^2}{\mu}\right)^{3/2}+\right.\\
&\left. +\frac{6L_{_{PL}}^2}{\mu}\left(x-\frac{2L_{_{PL}}^2}{\mu}\right)^{1/2}\right).
\end{split}
\label{solution-20}
\end{equation}
Once again, (\ref{solution-20}) has no singularities ($x_{{\rm min}}=2L_{_{PL}}^2/\mu$), and at $2L_{_{PL}}=0$ reduces to a familiar solution: $x\propto t^{2/3}$.

\item A universe with a cosmological constant $\rho=\pm\lambda^2$. For the positive cosmological constant the general solution of \eqref{con2} is:
\begin{equation}
\begin{split}
x(t) &= \frac{1}{\sqrt{2}\lambda}\sqrt{k c^2 + \gamma_{_+} \cosh\left(2\lambda(t-t_0)\right)}, \\
\gamma_{_+} &=\sqrt{c^4+\left(4 c \lambda L_{_{PL}}\right)^2},
\end{split}
\label{dS}
\end{equation}
and it is nonsingular, well-defined for all $k=0,\,\pm 1$, and at $L_{_{PL}}\to 0$ converges to classic ''dS'' solutions. The case of a negatively-defined $\rho$ is trickier, since it only allows for the real-valued solutions when the universe is open. The solution itself then has the form:
\begin{equation}
\begin{split}
x(t) & = \frac{1}{\sqrt{2}\lambda}\sqrt{c^2 + \gamma_{_-} \sin\left(2\lambda(t-t_0)\right)},\\
\gamma_{_-} &= \sqrt{c^4 - \left(4 c \lambda L_{_{PL}}\right)^2}.
\end{split}
\label{AdS}
\end{equation}
The formula (\ref{AdS}) is a generalization of the familiar AdS solution (and converges to it at $L_{_{PL}}=0$), with the quantum potential radically altering the normal outcome to produce a nonsingular oscillating universe with a rebound similar to (\ref{solution-2pl}).

\item Now let us consider the special case of (\ref{sing}), when $w=1/3$, $\rho(x)=R^2/x^4$. In other words, let us take a look at the radiation-dominated universe. The behaviour of the model will depend on the quantity
\begin{equation}
\delta=4\left(\frac{2\pi GR^2}{3}-L_{_{PL}}^2c^2\right),
\label{delta}
\end{equation}
producing three alternatives:

{\bf (i)} $\delta>0$. In this case the solution will be rather similar to a classical model of a radiation-dominated universe:
\begin{equation}
x(t)=c\sqrt{kt\left(\frac{2\sqrt{\delta}}{c^2}-t\right)},\qquad {\rm if}\,\,k^2=1,
\label{R1}
\end{equation}
or
\begin{equation}
x(t)=\sqrt{2\sqrt{\delta}t},\qquad {\rm if}\,\,k=0.
\label{R2}
\end{equation}
The constants of integration in (\ref{R1}), (\ref{R2}) are so chosen to satisfy the initial condition $x(0)=0$. The open universe ($k=+1$) emerges from the singularity (at $t=0$) and ends in one at $t=2\sqrt{\delta}/c^2 = T_{\rm {life}}$. The total lifetime $T_{\rm {life}}$ of such a universe differs from the lifetime $T_{\rm {clas}}$ of a purely classical universe (which it converges to as $L_{_{PL}}\to 0$) by a tiny margin, measured in Planck time:
\beqn
\begin{split}
T_{\rm {life}} &=\sqrt{T_{\rm {clas}}^2-4\tau_{_{PL}}^2 }, \\
\tau_{_{PL}} &=\frac{L_{_{PL}}}{c}=10^{-43}\,\,{\rm sec}.
\end{split}
\enqn
Interestingly, this implies that, quite unlike the dust dominated universe (\ref{solution-2pl}), the period of oscillation now actually depend upon the quantum effects (albeit in an admittedly almost imperceptible manner). The solution (\ref{R1}) with $k=-1$ describes an open universe emerging from an initial singularity at $t=2\sqrt{\delta}/c^2$, and (\ref{R2}) deals with a flat universe, which is also born from a singularity at $t=0$.

{\bf (ii)} $\delta=0$. This case as well the next one can only be realized in the presence of a quantum potential. The equation (\ref{con2}) reduces to ${\dot x}^2+kc^2=0$ which has real-valued solutions either when $k=-1$ ($x(t)=\pm c(t-t_0)$) or $k=0$ ($x={\rm const}$).

{\bf (iii)} $\delta<0$. The solution of \eqref{con2} is very similar to a pure quantum case (\ref{solution-1}):
\begin{equation}
x(t)=\sqrt{c^2\left(t-t_0\right)^2+\frac{|\delta|}{c^2}},
\label{R3}
\end{equation}
and the universe must be necessarily open ($k=-1$).

\item For our last example of a two-universe ensemble let us consider a second universe filled by a matter with a stiff equation of state $w=1$ and density $\phi(x)=B^2/x^6$. The equation (\ref{con2}) is again completely integrable, but the general solutions become rather cumbersome. For this reason we will restrict ourselves to providing only the solution for a simplest case of a flat universe:

\begin{equation}
\small{
\begin{split}
\arcsin\left(\frac{L_{_{PL}} x}{m}\right)&-\frac{L_{_{PL}} x}{m}\sqrt{1-\left(\frac{L_{_{PL}} x}{m}\right)^2}=\\
&=\pm \frac{4cL_{_{PL}}^3}{m^2}\left(t-t_0\right),
\end{split}
}
\label{pred}
\end{equation}
where $m=(B/c)\sqrt{2\pi G/3}$. This is a singular solution, since $x(t_0)=0$, which lies in accordance with the condition (\ref{sing}). Expanding (\ref{pred}) into the Taylor series as $L_{_{PL}}\to 0$,
it is possible to show that the classical limit produces a well-known formula $x\propto (t-t_0)^{1/3}$.

\end{itemize}

\subsection{Three is a company: $N=3$  and the Hamilton-Jacobi equation}\label{ssec:N3}

Let us broaden our outlook and consider an ensemble of {\em three} universes: a Barrow's  zero universe with $a_1=0$, a purely ``quantum universe'' with scale factor $a_2$ and a third universe with scale factor $a_3$, filled by the fields of matter that satisfy the condition \ref{condition4}. Of course, we assume the scale factors $a_1, a_2, a_3$ to satisfy the condition \ref{condition1}, so that the Theorem \ref{Theorem1} ensures there would be no singularities. The quantum potential (\ref{U}) for our new multiverse takes the form:
\begin{equation}
U=\frac{1}{a_2^2}+\left(\frac{1}{a_3-a_2}-\frac{1}{a_2}\right)^2+\frac{1}{(a_3-a_2)^2}.
\label{N3}
\end{equation}
For what follows it will be handy to replace the scale factors with new radial field variables: $\{a_2(t),\,a_3(t)\}\to \{r(t),\,\phi(t)\}$ as follows:
\begin{equation}
a_3+ia_2=r{\rm e}^{i\phi},\qquad 0<\phi<\frac{\pi}{4}.
\label{rad}
\end{equation}
Now, what kind of fields of matter should fill the third universe? In this subsection we will consider three cases: (I) there are no fields of matter ($\rho_3=p_3=0$) and the third universe is, like the second one, governed purely by a quantum potential; (II) a universe filled with matter whose parameter of state is $w=-1/3$ (and so $\rho_3=\sigma^2/a_3^2$); (III) a hot universe filled by radiation with $w=1/3$ (here $\rho_3=R^2/a_3^4$). The reason we chose them specifically is because in these three cases the equations turns out to be completely integrable and leading to solutions with a strikingly similar behaviour. It is crucial to keep in mind that unlike the previously discussed cases, when $N>2$ the equations \eqref{LE}, \eqref{con} generally cease to be completely integrable, so the importance of those rare cases with known exact solutions cannot be overestimated. In fact, in the remainder of this section we will discuss both the solutions and the method of obtaining them.

In all three cases, after switching to the new variables (\ref{rad}), the equation (\ref{con}) turns into:
\begin{equation}
\frac{1}{2}\left({\dot r}^2+r^2{\dot\phi}^2\right)+\frac{v(\phi)}{r^2}=E,
\label{rad1}
\end{equation}
where the shape of function $v(\phi)$ is case-specific: for the cases (I) and (II)
\begin{equation}
\displaystyle{
v(\phi)=\frac{2c^2 L_{_{PL}}^2(4-2\cos 2\phi-3\sin 2\phi)}{(1-\cos 2\phi)(1-\sin 2\phi)},}
\label{vI-II}
\end{equation}
and case (III) requires
\begin{equation}
\begin{split}
v(\phi)=&\frac{16 \pi G \hbar}{3c}\left(\frac{4-2\cos 2\phi-3\sin 2\phi}{(1-\cos 2\phi)(1-\sin 2\phi)}\right)-\\
&-\frac{8 \pi G R^2}{3(\cos 2\phi +1)}.
\end{split}
\label{vI-II}
\end{equation}
Similarly, $E=c^2(1-k_3)/2$ for (I) and (III), but
$$
E=\frac{4\pi G\sigma^2}{3}-\frac{c^2}{2}\left(k_3-1\right),
$$
for case (II). Note that $k_2=-1$, whereas $k_3=-1$ for the case (I) and $k_3=0,\,\pm 1$  for (II) and (III).

In order to integrate (\ref{rad1}) we will utilize the Hamilton-Jacobi approach with the following sequence of steps:

1. In the beginning, the generalized momenta $p_r={\dot r}$, $p_{\phi}=r^2{\dot\phi}$ are introduced so that the l.h.s. of (\ref{rad1}) might be rewritten as a Hamilton function: $H(p_r,\,p_{\phi},\,r,\phi).$

2. Then we introduce an action function $S=S(t,r,\phi)$ which allows us to replace the generalized momenta with the corresponding derivatives: $p_r\to {\partial S}/{\partial r}$, $p_{\phi}\to {\partial S}/{\partial\phi}$. As a result, (\ref{rad1}) turns into an abbreviated Hamilton-Jacobi equation for conservative systems:
\begin{equation}
\displaystyle{
\frac{1}{2}\left(\frac{\partial S}{\partial r}\right)^2+\frac{1}{2r^2}\left(\frac{\partial S}{\partial \phi}\right)^2+\frac{v(\phi)}{r^2}=E.}
\label{311}
\end{equation}

3. On this step we find the complete integral of (\ref{311}), which depends upon two arbitrary constants: $E$ and $\alpha$~\footnote{In fact, there are three constants, but the third one is only important in the construction of the general integral.}:
\begin{equation}
S=S(t,r,\phi,E,\alpha)=-Et+f(r,\phi,E,\alpha).
\label{312}
\end{equation}

4. Next, we differentiate (\ref{312}) w.r.t. $E$ and $\alpha$ and equates= the results to new constants $\beta_{1,2}$:
\begin{equation}
\frac{\partial}{\partial E}S(t,r,\phi,E,\alpha)=\beta_1,\qquad \frac{\partial}{\partial \alpha}S(t,r,\phi,E,\alpha)=\beta_2.
\label{313}
\end{equation}

5. We use \eqref{313} to derive the sought after quantities $r$ and $\phi$ as functions of $t$, $E$, $\alpha$ and $\beta$. The required solution can be constructed out of them by simply fixing $E$.

It must be said that, in general, finding a complete integral of the Hamilton-Jacobi equation is a tremendously difficult task. Fortunately, our case is an exception, because the equation (\ref{311}) is actually separable, its complete integral having the form
\begin{equation}
\begin{array}{l}
\displaystyle{
S(t,r,\phi,\alpha,E)=-Et+\kappa_2\sqrt{2}\int d\phi\sqrt{\alpha-v(\phi)}+
}\\
\\
\displaystyle{+\kappa_1\sqrt{2\alpha}\left(\sqrt{\frac{Er^2}{\alpha}-1}-\arctan\sqrt{\frac{Er^2}{\alpha}-1}\right),
}
\end{array}
\label{314}
\end{equation}
where $\kappa_{1,2}^2=1$,  $\alpha>0$ and $E>0$  (we'll consider the special case $E=0$ at the end of this section). It immediately follows from (\ref{314}) that the function $r(t)$ ought to be similar in all three cases. Next, by taking a partial derivative of (\ref{314}) w.r.t. $E$ and resolving the resulting equation \eqref{313} for $r$ yields the following formula:
\begin{equation}
r(t)=\sqrt{a_2^2(t)+a_3^2(t)}=\sqrt{\frac{\alpha}{2E}+2E(t+\beta_1)^2},
\label{315}
\end{equation}
which states that for $t\to \infty$ we shall have $\sqrt{a_2^2+a_3^2}\to \sqrt{2E}t$, and, more importantly, that $r(t)\ne 0$ $\forall\,\,{t}$. In other words, we have just constructed the solution which is patently non-singular, in complete accordance with Theorem \ref{Theorem1}.

The second equation in \eqref{313} is obtained via the differentiation w.r.t. $\alpha$. Doing exactly that and using a newly derived relationship (\ref{315}) results in
\begin{equation}
\displaystyle{
\int\frac{d\phi}{\sqrt{\alpha-v(\phi)}}=2\beta_2+\frac{1}{\sqrt{\alpha}}\arctan\left[\frac{2 E}{\sqrt{\alpha}}\left(t+\beta_1\right)\right].}
\label{317}
\end{equation}
The behaviour of the solution \eqref{317} explicitly depends upon the choice of $v(\phi)$ and will be different in cases (I), (II) and (III). On the other hand, we can still surmise asymptotic behaviour of the universes $\# 2$ and $\# 3$ without actually taking the integrals in (\ref{317}). Indeed, using (\ref{rad}) we can calculate the Hubble parameter:
\begin{equation}
H_2=\frac{{\dot r}}{r}+\cot\phi\,{\dot\phi},\qquad H_3=\frac{{\dot r}}{r}-\tan\phi\,{\dot\phi}.
\label{319}
\end{equation}
Differentiation it w.r.t. $\phi$, expressing ${\dot\phi}$ out of it, and inserting the result into (\ref{319}) produces the following:
\begin{equation}
\begin{array}{l}
\displaystyle{
H_2=\frac{2E(2E(t+\beta_1)\tan\phi+\sqrt{\alpha-v(\phi)})}{\tan\phi\left(4E^2(t+\beta_1)^2+\alpha\right)},}\\
\\
\displaystyle{
H_3=\frac{2E(2E(t+\beta_1)-\tan\phi\sqrt{\alpha-v(\phi)})}{4E^2(t+\beta_1)^2+\alpha}.}
\end{array}
\label{320}
\end{equation}
Since $\tan\phi$ and $\sqrt{\alpha-v(\phi)}$ are bounded on the intervals under the consideration, as $t \to\infty$:
\begin{equation}
H_2\to  H_3\to \frac{1}{t}.
\label{H23}
\end{equation}

Thus, in time both universes overcome their difficulties by adopting asymptotically similar regimes of expansion. One might ask: is it only true for the specific cases we've chosen, or is it a more general property of MIU framework? A physical intuition tells us that it is the latter, but the proof would require a healthy dose of numerical modeling which lies out of scoop of this article.


One last case we should consider here, though, would be that of a closed ($k_3=+1$) third universe filled by radiation. Here $E=0$ and we must impose an additional condition $\alpha \propto E$. For simplicity, let the coefficient of proportionality be equal to 2. The solution of the resulting model will non-singular and describe two universes that oscillate with a phase shift $\pi/2$; their scale factors being equal to $a_2=\sin\phi(t)$ and $a_3=\cos\phi(t)$, where $\phi(t)$ is determined via

\begin{equation}
\displaystyle{
\int\frac{d\phi}{\sqrt{-v(\phi)}}=2\beta_2+\beta_2+t,}
\label{k3=1}
\end{equation}
and  $v(\phi)$ is given by (\ref{vI-II}). Note, that except for the pathologically small values of constant $R \to 0$, the function $-v(\phi)$ remains strictly positive in between the stop points $\phi_1$ and $\phi_2$ that satisfy the condition $0<\phi_1<\phi_2<\pi/4$, thus making sure the solutions (\ref{k3=1}) remain real valued.

The examples we have discussed clearly demonstrates the effect of Theorem \ref{Theorem1}. Furthermore, the solutions are clearly perturbative w.r.t. $\hbar$. In fact, from the point of view of peturbativity the most interesting case would be the one of an ensemble of $N$ universes, consisting of a Barrow's zero universe, a purely quantum universe and the rest being the closed and radiation-dominated universes with various scale factors. In such a multiverse the equations (\ref{LE}) and (\ref{ner}) reduce to
\begin{equation}
\sum_{n=2}^N a_n^2=At+B,
\label{deco}
\end{equation}
where $A$ and $B$ are two constants of integration. The condition (\ref{con}) imposes a severe restriction $A=0$ (see the previously discussed example of a closed universe), but it no longer applies in the classical limit $\hbar=0$, thus producing the well-known classical solutions. For example, if $N=3$ and $\hbar =0$ the expression (\ref{deco}) turns into $a_2^2+a_3^2=c^2Tt$ ($T={\rm const}$), where $B$ is eliminated by a simple time translation and $A=c^2T$.  Because of (\ref{clas}) we must have $a_2=ct$. The solution $a_3$ immediately takes a familiar form of a classical Friedman solution of a closed radiation-dominated universe:
$$
a_3(t)=c\sqrt{Tt\left(1-\frac{t}{T}\right)}.
$$

\section{A MIU to a kill: how Many Interacting Universes model disposes of the Big Rip singularities}\label{Sec:Phantom}

In this Section we are going to step outside of the confines of weak energy condition-compliant cosmologies and prove a rather startling proposition: that Theorem \ref{Theorem1}, being so effective in ridding of the $a=0$ type singularities, is in fact equally effective in exterminating the Big Rip singularites $a=\infty$. Such multi-functionality immediately implies two things: first, that MIU quantization is compatible with the phantom types of matter (provided they actually exist), and secondly, it opens up an exciting possibility of a future quantum gravity theory which might be completely free from {\em any type} of singularities, be it with a zero or a infinite values of a scale factor!

\subsection{A Phantom Nemesis?}\label{Sec:Phan}

It is presently unknown whether the observed accelerated expansion of our Universe can be generated by some fundamental phantom fields. Even a question of whether such fields might exist at all is a subject of rather heated debates  \cite{2.1}, \cite{2.2}, \cite{2.3}, \cite{2.4}, \cite{2.5}. What is relevant for us here is one undeniable fact: the Big Rip singularities, deeply associated with such phantom fields, are utterly unreconcilable with the very method of quantization via MIU. To understand why, let us first recall that the thermodynamics of the phantoms require them to possess {\em negative} absolute temperatures \cite{PedroCarmen},  see also \cite{LimaAlcaniz}. Consider, for example, a universe with scale factor $a$, filled by a minimally coupled scalar field $\phi$. A density of the kinetic energy is proportional to a density of temperature $T$, so
\begin{equation}
T \propto a^3{\dot\phi}^2=a^3(w+1)\rho=K(w+1)a^{-3w},
\label{Tem}
\end{equation}
where $K>0$ is a constant of integration. A phantom field has $w<-1$, so the temperature (\ref{Tem}) ought to be negative. On the other hand, an entropy of a phantom field can be shown to remain positive \cite{PedroCarmen}~\footnote{One can easily verify this by expressing $\rho=\rho_0a^{-3(w+1)}$ via the temperature (\ref{Tem}) and then using the Second law of thermodynamics.}. The most natural explanation of this discrepancy must lie in a fundamentally quantum mechanical nature of the phantoms. And, since arguably the strongest manifestation of a phantom field is a Big Rip singularity, it in turn must also be a purely quantum phenomenon.
An alternative way to reach this conclusion lies in observing the growth of an energy density at a vicinity of Big Rip: once it reaches the Planck value $\rho_{_{PL}}=c^5\hbar^{-1}G^{-2}$, a correct description of the dynamics can no longer be classical and has to account for the quantum phenomena. Anyway, one way or another, we have apparently stumbled upon a serious problem. According to the MIU formalism, all quantum mechanical effects have but one source: a repelling quantum potential (\ref{U}). And this potential will obviously vanish during the uninhibited growth of the scale factor as we approach the Big Rip singularity! Hence, according to the MIU model, the {\em quantum mechanical effects will vanish} just as we approach Big Rip. So, we end up with a nasty contradiction on our hands, which can only be resolved in two possible ways: (1) The MIU approach might be incorrect (needless to say, this would be a huge blow for the authors!) or, at the very least, it might require certain modifications to ensure the growth of the quantum effects near the Big Rip; (2) The MIU is correct, and the problem is alleviated because {\em the quantum potential actively prevents the Big Rip singularities from arising} just like it did with big bang/crunch singularities. Very fortunately for the authors, the very conditions we have used in Theorem \ref{Theorem1} ensures that the latter is actually true!

Before we proceed to the crux of our proof, let us list everything we'll need and will use. First of all, for simplicity we will stick to a unit system where $8\pi G/3=c=1$. Secondly, we will use the equation (\ref{LE}), (\ref{con}) with the potential (\ref{U}) and a boundary condition (\ref{bc}). Thirdly, the continuity condition (\ref{ner}) must also be taken into account. Finally, out of four conditions from Sec. \ref{Sec:MIU} we will utilize the first three, (\ref{C1}), (\ref{C2}), (\ref{C3}), but not the fourth one -- after all, one can hardly hope to vanquish the Big Rip singularities if they are not even given a chance to appear!

Now properly armed, we can formulate the main theorem of this Section:

\begin{theorem} \label{Theorem2}
If the conditions \ref{condition1}-\ref{condition3} are satisfied, the solutions of the system (\ref{LE}) and (\ref{con}) suffer no Big Rip singularities for all $n > 1$.
\end{theorem}

We will prove this theorem in (\ref{Sec:NoBR}), but before we go, we would like to make one last observation. Let us assume for a second that Theorem \ref{Theorem2} is incorrect and at least one of the universes of number $n$ experiences a Big Rip at certain time $t_s$. Then it immediately follows that the neighbouring universe with number $(n+1)$ must also undergo the Big Rip at some $t\le t_s$. Indeed, the repulsive quantum force between the $n$'s and $(n+1)$'st universes shall either stop the unlimited growth of $a_n$, thus negating the BR-singularity and proving the Theorem \ref{Theorem2}, or fail to do so, forcing $a_{n+1}$ to grow at least as fast as $a_n$, because otherwise there will exist such  $t'<\infty$, when $a_n(t')=a_{n+1}(t')<\infty$, forcing the potential (\ref{U}) to diverge, and totally violating condition (\ref{con}) in the process. Hence, the bigger the number the universe has, the earlier it will experience the Big Rip (if at all!). In other words, for a finite ensemble of $N$ interacting universes it would be sufficient to prove Theorem \ref{Theorem2} for the largest unverse of number $N$. Let us now get on with it!~\footnote{Note that the finiteness of number $N$ significantly simplifies the proof of Theorem \ref{Theorem2}, but is actually not necessary and can be omitted.}

\subsection{The Step-by-Step Proof of Theorem \ref{Theorem2}}\label{Sec:NoBR}

Our proof will be done in the classical {\em reductio ad absurdum} fashion: we will see how the assumption that the universe number $N$ (the largest n the ensemble) suffers a Big Rip at some time $t_s$ naturally leads to a mathematical contradiction. The key to the proof of Theorem \ref{Theorem2} is a curious mathematical technique previously used by one of the authors to discover a phenomenon of the phantom zone crossing \cite{BRRaz}, and then explored in more details in \cite{Neopr}. To avoid unnecessary confusion we have deemed it wise to break the proof into five logical steps, beginning with...

\begin{enumerate}[{\bf Step 1.}]
\item Let's additionally assume that we have managed to find the exact solutions of (\ref{LE}), (\ref{con}), i.e. that there exist $(N-1)$ time-dependent functions $a_n(t)$ ($n=2,\,3,...,\,N$) and $2(N-2)$ time-dependent functions $\rho_n(t)$, $p_n(t)$, with $n=3,...,\,N$ (recall that the universe with $n=1$ is a Barrow's zero-universe (\ref{ZU-gen}), whereas the universe with $n=2$ is purely quantum and has no fields of matter). This means that we can define $N-2$ ($n\ge 3$) time-dependent functions $v_n(t)$:
\begin{equation}
v_n(t)\equiv -\frac{1}{2}\left(\rho_n(t)+3 p_n(t)\right)-\frac{\hbar}{a_n(t)}\frac{\partial}{\partial a_n}U(a_1,...,a_{_N}),
\label{vn(t)}
\end{equation}
and formally write the system $N-1$ equations (\ref{LE}) as:
\begin{equation}
\begin{split}
{\ddot a_2}(t) &=-\hbar\frac{\partial}{\partial a_2}U(a_1,...,a_{_N}),\\
{\ddot a_n}(t) &=v_n(t)a_n(t),\qquad n\ge 3.
\label{sys}
\end{split}
\end{equation}

\item Since $N-2$ functions (\ref{vn(t)}) are known, one can derive $N-2$ new functions, henceforth known as ${\hat a}_n(t)$, such that the for all $n\ge 3$ the Wronskians $W(a_n(t),\,{\hat a}_n(t))=1$. Naturally, the functions will have the form
\begin{equation}
{\hat a}_n(t)=a_n(t)\int\frac{dt}{a^2_n(t)},\,\,\,n\ge 3,
\label{hata}
\end{equation}
and they will also be solutions of Sch\"odinger equations (\ref{sys}) with the same ``potentials'' $v_n(t)$, but {\em linearly independent of} $a_n(t)$\footnote{Generally speaking, if $y_1(x)$ is a non-trivial partial solution of ordinary linear O.D.E. $y'' + p(x) y' + q(x) y = 0$ where $p(x)$ is a continuous function, then the linearly independent solution $y_2(x)$ of the same equation will have a form:
\beqn
y_2 = y_1(x) \int{\frac{\eb^{-\int{p(x)dx}}}{y_1^2(x)} dx}.
\enqn
It is easy to check that in this case the Wronskian $W(y_1, y_2)=1$}.

\item What about ${\hat a}_2(t)$, which does not satisfy a Schr\"odinger equation? In order to find it we can substitute $N-2$ newly found functions ${\hat a}_n(t)$ plus the hitherto unknown ${\hat a}_2(t)$ into (\ref{U}). The result will be the following nonlinear O.D.E. of second order:
\begin{equation}
\frac{d^2 {\hat a}_2(t)}{dt^2}=-\hbar\frac{\partial}{\partial {\hat a}_2}U({\hat a}_2,...,{\hat a}_{_N}),
\label{hata2}
\end{equation}
By solving \eqref{hata2} for ${\hat a}_2(t)$ we would successfully define $N-1$ new functions ${\hat a}_n(t)$ for $n\ge 2$ -- the last function ${\hat a}_1 = a_1(t)$ we'll assume to be the same zero universe (\ref{ZU-gen})). Interestingly, the initial ordering of original scale factors $a_n(t)<a_{n+1}(t)$ $\forall\, n$ will generally be inverted for the ``crowned'' scale factors -- i.e. we will usually have ${\hat a}_n(t)>{\hat a}_{n+1}(t)$ for $n>2$.

\item The $N-1$ functions ${\hat a}_n(t)$, $n\ge 2$, found on Step 3 will be the solutions of the same equations (\ref{LE}), (\ref{con}) (as well as the system (\ref{sys}) with $a_n\to {\hat a}_n$ and the original ``potentials'' $v_n(t)$), but with the {\em different fields of matter}. In other words, they will correspond to $N-2$ new densities ${\hat\rho}_n(t)$ and new pressures ${\hat p}_n(t)$, whose forms can be derived as follows.
    \begin{enumerate}
    \item     First, in the definitions of (\ref{vn(t)}) we shall replace all $a_n(t)$ by ${\hat a}_n(t)$ ($n=2,..,N$), and all $\rho_n(t)$, $p_n(t)$ by the unknown functions ${\hat\rho}_n(t)$ and ${\hat p}_n(t)$ ($n=3,..,N$).

    \item The way $v_n(t)$ depend on $t$ does not change, so the l.h.s. of (\ref{vn(t)}) is known, and we can therefore derive ${\hat p}_n(t)$ out of them.

    \item These ${\hat p}_n(t)$ we can then substitute into $N-2$ continuity equations -- thus guaranteeing that (\ref{con}) with newly ``crowned'' ${\hat a}_n$, ${\hat \rho}_n$ and ${\hat p}_n$ remains valid.

    \item Integrating the resulting $N-2$ linear O.D.E.s of first order, we can find ${\hat\rho}_n(t)$. Substituting them back into the continuity equations \eqref{ner}, we also find ${\hat p}_n(t)$. Note, that it will suffice to solve just $N-3$ equations, since ${\hat\rho}_3(t)$ is then easily derived from (\ref{con}).
    \end{enumerate}

    Note, that although one can hardly hope to perform the aforementioned steps analytically (the total amount of equations is most likely to be of an order of googleplex), the procedure itself is very simple and straightforward. Another important point is that we can completely omit this Step for the purely quantum universe with $n=2$ -- after all, its new scale factor ${\hat a}_2(t)$ have already been discovered at Step 4, and  it has no fields of matter to worry about. To sum everything up, at Step 5 we construct a complete set of new solutions of (\ref{LE}), (\ref{con}). Their physical nature should not concern us here -- only the fact that $a_n(t)$ and ${\hat a}_n$ are linearly independent partners of the same linear equations \eqref{sys}.

\item According to Theorem \ref{Theorem1} which we have proved in Sec. \ref{Sec:no-singularity}, the scale factors $a_n(t)$ are nonsingular for all $n>1$, which means that all ${\hat a}_n(t)$ defined by (\ref{hata}) are nonsingular for all $n>2$. In fact, ${\hat a}_2(t)$ is also nonsingular, which follows directly from (\ref{hata2}). Thus, the Theorem \ref{Theorem1} is fully applicable to ${\hat a}_n(t)$.  However, let us now recall that by assumption $a_{_N}(t)$ experiences a Big Rip singularity: $a_{_N}(t)\to\infty$ as $t\to t_s<\infty$. It follows from (\ref{hata}) that
\begin{equation}
\lim_{t\to t_s}\frac{d}{dt}\left(\frac{{\hat a}_{_N}(t)}{a_{_N}(t)}\right)=\lim_{t\to t_s}\frac{1}{a^2_{_N}(t)}=0.
\label{dokaz}
\end{equation}
Hence, as $t\to t_s$, ${\hat a}_{_N}(t)\to C a_{_N}(t)$, where $C={\rm const}$. But then the Wronskian
\begin{equation}
\lim_{t\to t_s} W(a_{_N}(t),\,{\hat a}_{_N}(t))\to 0,
\label{W01}
\end{equation}
which {\em contradicts the linear independence condition} between $a_{_N}(t)$ and ${\hat a}_{_N}(t)$. Hence, $C=0$ and ${\hat a}_{_N}(t)\to 0$ when $t\to t_s$. In other words, ${\hat a}_{_N}(t_s)=0$, which in turn contradicts the Theorem \ref{Theorem1}. Thus, the initial assumption about the existence of a solution (\ref{LE}), (\ref{con}) $a_{_N}(t)$ with a Big Rip singularity is false. Q.E.D.
\end{enumerate}

\section{Anisotropy, the Decoherence and the Initial Conditions Problem}\label{Sec:discussion}

One can safely say that the construction of successful quantum theory of gravity to this day remains one of the toughest problems in theoretical physics. Most of the efforts in this field are concentrated on two competing approaches of string theory and loop gravity~\footnote{The sheer quantity of works on both subjects is so overwhelming that it would be a folly to even attempt providing any kind of literary overview here.}. The approach that we have developed in this article is different in this regard, mostly because it is designed with a relatively smaller goal in mind: to apply the latest interpretation of the quantum mechanics (the MIW formalism, presented in \cite{MIW}) for the gravity models in hope that the novel approach might help to uncover some previously unknown relationships, indistinguishable by the complex mathematical apparatus of either string theory or loop gravity. The first attempt at this has been made at \cite{YY}, where we introduced the MIU formalism (a cosmological generalization of MIW) and explored its predictions using the master-factor method. In the present article we are looking at a more general model, with an important additional ingredient in the form of very special ``ill-posed'' solutions, recently discussed at length by Prof. Barrow in \cite{Zero-Univ}. The result is nothing short of remarkable: if one takes the Barrow's solutions into account and uses the MIU formalism with a few simple and realistically-sounding assumptions, then the long-maligned problem of gravitational singularities simply evaporates. In other words, the quantum gravity effects might produce a bouncing effect, stopping the gravitational collapse before the singularity is formed. For a next step, one might try incorporating the idea of ``zero universes'' into the more traditional approaches to quantization. However, even keeping in mind the old maxima that the ``looks can be deceiving'', it nevertheless {\em looks} to the authors that even by itself the MIU approach can prove to be useful for the task of quantization of gravity, and as such allows for further development and generalizations.

For example, we were so far restricting ourselves with the symmetric, homogeneous and isotropic cosmological solutions (the Friedman-Lema\^itre-Robertson-Walker models), mostly because generalizing the MIW formalism \cite{MIW} for these models is exceptionally easy and straightforward \cite{YY}. On the other hand, it is well known that the collapsing universe approaches the final singularity in an oscillatory manner, which is neither homogeneous nor isotropic. It consists of a sequence of so called Kasner epochs \cite{Kas}, \cite{Bel}, the order of which is characteristically chaotic~\footnote{The chaotic nature of a universe collapsing into the final singularity has been studied in huge volume of articles; it particular, it has been shown that in order to understand the dynamics one has to borrow from such disparate areas of mathematics as the theory of continuous fractions, the hyperbolic Kac-Moody algebras and the superstring theory in 10-dimensional space-time continuum. We will excuse ourselves from not citing any of those works because their subject lies well outside out of scope of the current article. Instead, we will refer our readers to a great review of the topic made by Kamenshchik \cite{Kamen}, with an excellent list of references}. But would it be so if we take into account the quantum potential? The answer requires a upgrading the MIU formalism for the more general cosmological models. For starters, one might consider the flat homogeneous anisotropic models with the metric
\begin{equation}
ds^2=dt^2-a^2(t)dx^2-b^2(t)dy^2-c^2(t)dz^2.
\label{aniz}
\end{equation}
Since this article is primarily concerned with the isotropic solutions, we will abstain from writing the complete Einstein equations for general anisotropic models. Instead we will provide one equation for a flat universe (for the sake of simplicity let us use the dimensional system where $8\pi G/3=c=1$), where $\rho$ is a zeroth component of the stress-energy tensor:
\begin{equation}
c\left(\dot{a}\dot{b}-\rho ab\right)+b\left(\dot{a}\dot{c}-\rho ac\right)+a\left(\dot{b}\dot{c}-\rho bc\right)=0.
\label{an-con}
\end{equation}
This equation is well-suited for building up the generalized constraint conditions akin to (\ref{con}). To do that one should first introduce $N$ universes with the metric (\ref{aniz}) and coefficients $a_n(t)$, $b_n(t)$, $c_n(t)$, $n=1,...,N$ and then replace (\ref{an-con}) with a sum over all those universes with an additional quantum potential $U(\{a_1,b_1,c_1\},..,\{a_{_N},b_{_N},c_{_N}\})$. The key question, of course, would the explicit form of this potential. For the homogenous and isotropic models we have used the potential in the form (\ref{U}), originally derived in \cite{MIW} for a one-dimensional problem. It seems logical to assume that the anisotropic potential can also be derived as a three-dimensional generalization of the potential from \cite{MIW}. Naturally, it would be premature to draw any serious conclusion without knowing its explicit form, but we do know that it has to be positively defined. And it was this very property that led us to a conclusion about the absence of singularities in FLRW models. Hence, we have a serious possibility that the quantization by the MIU model equipped with the Barrow's zero-universe might get rid of singularities even in anisotropic universes.

But the benefits of MIU formalism are not limited to the problem of existence of singularities. In our opinion, it might also shed light upon some other fundamental problems abound in the cosmology of the early universe, such as the problem of decoherence in the eternal inflation theory, and the problem of initial conditions for engaging with the inflation. Let us discuss them, beginning with the former.

The decoherence problem has been most coherently described in  \cite{Susk-B} and here is the crux of it. It is known that the eternal inflation describes the process of permanent creation of pocket universes (a.k.a. vacuum bubbles) in the same way the overheated water keeps producing the bubbles of water vapors. The description of this process is based on the well developed theory of classical stochastic processes. But in order for that theory to be applicable at all, first a decoherence must befall the original quantum superposition of vacuum decays -- and we remind our readers that these decays occur in different places at different times. If there is no decoherence there would be no classical bubble universes. But the problem is, as shown in \cite{Susk-B}, such a globally defined decoherence is {\em impossible}, and this spells very nasty (albeit not lethal, at least according to Susskind and Bousso) troubles for the eternal inflation theory. For more details on this problem we refer the reader to the aforementioned article by Bousso and Susskind. Here we would just like to point out that the problem of decoherence completely disappears in the MIU approach for an obvious reason -- the framework of MIU require neither wave functions nor density matrices, and therefore, no decoherence. In fact, what we call a ``decoherence'' in quantum mechanics from the point of view of MIU is merely a vanishing quantum potential. And this, in turn, implies that the dialect of MIU potentially can be more successful in describing the early universe than the standard language of quantum theory.

A second major problem to ponder is the problem of proper initial conditions for the inflation. The theory of inflation is truly majestic for a number of reasons, not the least of whom is an readiness with which it explains the thermodynamical arrow of time. Unfortunately, the is a tiny cloud that brings an unwelcome shade upon the majesty -- namely, the initial conditions required for the inflation must have radically low entropy \cite{Entropy1}, \cite{Entropy2}. It is commonly believed that the solution for the problem lay dormant somewhere in the future theory of quantum gravity. Now, we would be first to fully admit that the MIU formalism is by no means that theory -- it is at best a few introductory accords of the upcoming grand symphony of quantum gravity theory. Nevertheless, we can still try and use the MIU technique as a sort of a sonar to at least try and adumbrate the contours of the future theory. In particular, we can ask ourselves: does this approach bring anything new to the problem of the initial conditions? We'd argue that it does!

Let us consider what we already know. We have analyzed the quantum potential produced by the interacting Friedman universes, and we have even discussed its extension to the anisotropic models. Of course, any realistic quantum potential shall include the contributions from all possible universes. But what kind of them will dominate at the early stages? To answer this, we have to keep in mind that we are discussing the early universes, where the quantum phenomena are still prevalent -- i.e. those universes with a significantly strong quantum potential. And what kind of universes with a given (small) values of scale factors feels the strongest pull of quantum potential? It seems that the Friedman universes do, because their MIU potential diverges as $a^{-2}$, whereas the preliminary studies show the anisotropic potential diverging as $(ab)^{-1}$, where $a$ contract but $b$ expands. This discrepancy might indicate that at Plank lengths most universes has no choice but be homogeneous and isotropic, simply to be consistent with their quantum mechanical description as newly formed bubbles of metastable vacuum. Granted, this is just a conjecture, but a tempting one to pursue! If correct, it would mean that the upcoming theory of quantum gravity might prove that the initial conditions to initiate the inflation are indeed both typical and ubiquitous, fulfilling the hopes of most cosmologists after all.

But whatever the outcome, at this stage we can safely say this: the ``zero universes'' might have zero scale factors, but as a concept they are much grander in scale and seem to factor in more cosmological phenomena then we believed so far. In fact, it is quite possible that Barrow has proposed a crucial piece of the puzzle, a key ingredient of a theory that would lead to a complete eradication of singularities in quantum cosmology. And one can't help but wonder what John would have said about it if he was still with us.
\newline
\newline
\appendixtitleon
\appendixtitletocon
\begin{appendices}
\section{From de Broglie-Bohm interpretation to the Cosmology of Many Interacting Universes} \label{sec:Appendix}

In this appendix we will explain the main idea behind and the derivation of the Many Interacting Universes concept. The good way to start would be at the very begging, which in this case is the concept of a ``pilot wave'', introduced by Louis de Broglie and David Bohm in the early 1950-s \cite{Bohm} as an alternative to a then prevalent Copenhagen interpretation of the quantum mechanics.

\subsection{Piloting the wave function: the de Broglie-Bohm approach}

Consider a classical spin-less particle of mass $m$ embedded in a potential $V(\vec q)$. According to quantum mechanical laws, the behaviour of this particle is predicated on the solution of the nonlinear Schr\"odinger equation
\beq \label{Schrodinger}
i \hbar \psi_t = -\frac{\hbar ^2}{2m} \Delta \psi + V \psi,
\enq
where the wave function $\psi(t,\vec q)$ serves a role of {\em probability amplitude} to discover our particle at a given state. As any complex-valued function, it might also be decomposed as:
\beq \label{psi_r_s}
\psi = r \eb^{i S/ \hbar},
\enq
where $r$ and $S$ are real-valued and satisfy the equations:
\beq \label{S_r}
\begin{cases}
\displaystyle{S_t = \frac{\hbar^2}{2m} \frac{\Delta r}{r} - \frac{(\nabla S)^2}{2m} - V} \\
\\
\displaystyle{r_t = - \frac{1}{m}\nabla r \cdot \nabla S - \frac{1}{2m} r \Delta S.}
\end{cases}
\enq

Decomposed or not, the wave function $\psi$ is a fundamental and physically meaningful quantity -- in fact, the Copenhagen interpretation states that prior to us observing said particle's position or momentum, none of them are physically meaningful -- but the wave function is! It is only through the act of observation, when the wave function ``collapses'' to a classical state, corresponding to a single ``classical'' term in $\psi$, that the particle can be said to exist at a certain place or possess a certain momentum.

The idea of the collapse of wave function as an indispensable part of quantum mechanics irked a lot a physicists: to them it looked like a blemish on the otherwise beautifully woven fabric of quantum formalism. One possible way to get rid of it was proposed in 1957 by Hugh Everett III in his doctoral dissertation \cite{MWI}. He argued that the different terms in the wave function describes actual physical universes, and that the observation simply randomly shafts the particle (and the observer) into one of them. In this approach there is no place for collapse, since every term in $\psi$ corresponds to its own alternative version of a particle, observed by its own alternative observer. In other words, from the point of view of this ``many-world interpretation'', a simple act of observation of quantum particle splits a single universe into a number of parallel universes, identical to each other in every respect but with different {\em outcomes} of said observation.

There was, however, another approach. In 1927, during the famous fifth Solvay Conference on Physics, the French physicist Louis de Broglie gave a talk entitled ``The New Dynamics of Quanta''. He surmised that, if the collapse of the wave function is not real, than the particles should possess definite positions and momenta not only {\em after} the observation, but at all times. In other words, the Hamiltonian dynamics must still be valid. We just need to account for some sort of additional {\em field} acting on those particles, which would be a cause for all known quantum phenomena. And, of course, the only physically meaningful candidate for such a field must be no other then the wave function itself, which serves as a sort of a ``pilot wave'' (as De Broglie has colorfully put it). This reasoning, later independently rediscovered by David Bohm, has since been dubbed the de Broglie-Bohm interpretation.

There are a number of ways to derive the correct form of the modified Newton equation that properly accounts for the quantum mechanical phenomena and is consistent with the Schr\"odinger equation. For this Appendix we will stick to a very simple derivation proposed by D\"urr, Goldstein and Zangh\'i in \cite{Durr}. First, let us assume that the field $\psi$ must indeed be directly responsible for altering the actual trajectories of a given quantum particle. This implies that there exists such a real-valued vector function $\vec v(\psi)$ that the velocity of said particle satisfies the relationship:
\beq \label{q_v}
\dot {\vec q} = \vec v(\psi).
\enq
What kind of function is $\vec v$? Its form can be glimpsed from our core assumption that $\vec v$ must be consistent with the Schr\"odinger equation \eqref{Schrodinger}. First of all, since \eqref{Schrodinger} is linear, it is invariant w.r.t. transformation $\psi \to c \psi$, where $c \in \C$. Hence, \eqref{q_v} must also be invariant, which implies that the function $\vec v$ must be {\em homogeneous of order zero}. The most natural candidate for such a function is
\beq \label{v-version_1}
\vec v_{candidate}(\psi) = \alpha \frac{\nabla \psi}{\psi}, \qquad \alpha = \text{const},
\enq
which has a natural benefit of turning to zero at local maxima of $\psi$, and diverging at simple zeros of $\psi$ -- the former ensures that the particle tend to remain at or near the maxima of the wave function, and the latter can help to explain why the particle tries to avoid the zeroes of $\psi$.

However, a careful observer might immediately notice something fishy about \eqref{v-version_1}. If we use the decomposition \eqref{psi_r_s}, it turns into
\beq
\label{v-version_2}
\vec v_{candidate}(\psi) = \alpha \frac{\nabla r}{r} + i \frac{\alpha}{\hbar} \nabla S,
\enq
which is quite absurd, since it is obviously complex-valued and, as such, cannot be a proper velocity. On the other hand, we can recall yet another property of the Schr\"odinger equation \eqref{Schrodinger}, namely that it is invariant w.r.t. the simultaneous time inversion $t \to -t$ and complex conjugation of its solution $\psi \to \psi^*$. Looking at \eqref{v-version_2}, we immediately realize that such a property indeed holds -- but only for the {\em imaginary} part of \eqref{v-version_2} with a likewise imaginary $\alpha$ \footnote{By the same reasoning we can also eliminate all possible terms of the form $(\nabla \psi / \psi)^n$ with $n \neq 1$}. Therefore, the proper form of real-valued $\vec v(\psi)$ must have a form
\beq
\label{v-version_3}
\vec v(\psi) = \frac{\beta}{\hbar} \nabla S, \qquad \beta \in \R.
\enq
Finally, let us elucidate the value of a new constant $\beta$. In order to do that, we shall utilize -- what else? -- yet another property of the Schr\"dinger equation: the Galilean invariance. It is well-known that during the transition to a new coordinate frame $\vec {q'}$ which moves with a velocity $\vec v_0$ w.r.t. to the original coordinate frame $\vec q$, the solution of \eqref{Schrodinger} gets ``boosted'':
\beqn
\psi(\vec {q'}) = \psi(\vec q) \cdot \eb^{- \frac{m}{\hbar} \vec v_0 \cdot \vec q} = r ~ \eb^{\frac{i}{\hbar}\left(S - m \vec v_0 \cdot \vec q\right)} = r \eb^{iS'/\hbar},
\enqn
If we require that the same rule apply to our velocity formula \eqref{v-version_3}, we will get:
\beqn
\vec v - \vec v_0 = \frac{\partial\vec {q'}}{\partial t} =  \frac{\beta}{\hbar} \nabla S' = \frac{\beta}{\hbar} \nabla S - \frac{\beta m \vec v_0}{\hbar},
\enqn
which can only be true if $\beta = \hbar / m$, at last providing us with the following important result: if the momentum $\vec p$ of a quantum particle is well-defined at all times, then it is intimately connected to the wave function via the formula:
\beq \label{p_S}
\vec p = m \dot {\vec q} = \nabla S.
\enq

Once we have \eqref{p_S}, it is then straightforward to find the corresponding ``quantum field'' that is supposed to propel the particle along its quantum mechanically consistent path: one just has to take a partial derivative w.r.t. $t$ of both sides of \eqref{p_S} and then use the first (Hamilton-Jacobi) equation of \eqref{S_r}:
\beq\label{p_U_V}
\dot{\vec p} = \nabla S_t = - \nabla\left(-\frac{\hbar^2}{2m} \frac{\Delta r}{r}\right) -\nabla V = -\nabla Q - \nabla V,
\enq
where we have used the fact that $p_j$ components of the momentum are independent of $q_j$. As expected, we end up with a new, additional potential $Q$:
\beq \label{Q_Bohm}
Q = -\frac{\hbar^2}{2m} \frac{\Delta r}{r}.
\enq
All the quantum mechanical phenomena are effectively stored within this repelling quantum potential $Q$, and if we take a semiclassical limit $\hbar \to 0$ the entire equation reverts back to a standard second Newton's law. Interestingly, $Q$ explicitly depends only upon $r$, which satisfies the continuity equation (see second eq. of \eqref{S_r}):
\beq
r_t + \nabla r \cdot \vec p = 0.
\enq
This equation, of course, ensures that $|\psi|^2=r^2$ can still serve as a probability density.

We can take one more step and extend this approach to multiple particles. For $K\ge 1$ particles consider a $D=3K$ dimensional space with vector $\vec q = \{q_j\}_{j=1}^{3K}$, where a $k$-th particle corresponds to a triplet of coordinates $\{q_{3k-2}, q_{3k-1}, q_{3k}\}$ and has a mass $M_k$, which can also be written as a triplet $M_k = m_{3k-2} = m_{3k-1} = m_{3k}$. The Schr\"odinger equation in a $D$-dimensional space takes a form
\beq \label{Schrodinger_D}
i \hbar \psi_t = -\sum\limits_{j=1}^{3K}\frac{\hbar ^2}{2 m_j} \frac{\partial^2 \psi}{\partial q_j^2} + V(\vec q) \psi,
\enq
and in effect describes a {\em world-particle} in a $D$-dimensional space with D-dimensional momentum $\vec p = \{m_j q_j\}_{j=1}^D$. It is easy to show that the same algorithm can be applied here, producing the $D$-dimensional analogue of \eqref{p_S}:
\beq \label{p_S_D}
\vec p = \sum\limits_{j=1}^{3K} \frac{\partial S}{\partial q_j},
\enq
as well as a generalization for the quantum potential $Q$:
\beq \label{Q_D_Bohm}
Q = -\sum\limits_{j=1}^{3K}\frac{\hbar ^2}{2 m_j r} \frac{\partial^2 r}{\partial q_j^2}.
\enq

To sum up, after 1950-s there existed two competing interpretations of quantum mechanics sans collapse of wave functions: the Many Worlds Interpretation (MWI), and the de Broglie-Bohm interpretation (dBB), with their own unique individual strengths: MWI having a very conceptually clear philosophy and dBB proving to be very effective in many types of calculations where the standard techniques revolving around the solutions of Schr\"odinger equation become too cumbersome to use (see, for example, \cite{Holland}, \cite{Licata}). And yet both theories have each had their own unique blemishes. Within a framework of DBB there was a debate over the status of equation \eqref{p_S} (or \eqref{p_S_D}) -- was it a fundamental law of nature (de Broglie's opinion), or merely an initial condition, which is overshadowed by the modified Newton's Law \eqref{p_U_V} (the view of Bohm)? On an even more serious note, the dBB-altered Newton's Law states that the behaviour and the energy of quantum particles depends upon a field $\psi$, {\em which is not generated by the particles themselves}, plus the field $\psi$ in question is supposed to evolve not locally (in a close vicinity of a given particle), but over the entire domain, including the regions where there never were any particles to begin with (the so-called ``empty branches'' problem) -- all of it in sharp contrast to the usual classical mechanical models.

In turn, the MWI has its own share of head-scratchers. Let us consider the wave function in a state of superposition:
\beqn
|\psi\rangle = \alpha_1 |\psi_1\rangle + \alpha_2 |\psi_2\rangle, \qquad |\alpha_1|^2+|\alpha_2|^2=1.
\enqn
The observation splits the particle into two, one with wave function $|\psi_1\rangle$, and one with $|\psi_2\rangle$. The probabilities that a given observable particle surfaces in the $k$-th world is equal to $|\alpha_k|^2$ (the Born rule). And if $|\alpha_1|^2=|\alpha_2|^2=1/2$, the MWI actually explains why: there are two outcomes, each corresponding to its own Everett world, and there are equal chances to end up in either of them. Everything is nice and simple, until we get to a situation where the probabilities are {\em not equal}: say, $1/3$ and $2/3$. What does this mean from the point of view of MWI? Does it mean that the first outcome takes place in one world whereas the second occurs in {\em two}? How many parallel worlds are we actually talking about? Perhaps we are dealing with a whole ensemble of worlds, and for every world graced by outcome No.1 there are twice as many worlds with outcome No.2? But how can it comply with our initial assumption that the Everett worlds are identical at the moment of the splitting in every respect except for the results of the measurement of a single particle?

All these problems led to the majority of physicists abandoning the notion of a ``proper'' interpretation of quantum mechanics, sticking instead to the famous mantra ``shut up and calculate''. However, in 2014 a group of theoretical scientists have proposed a completely new interpretation, combining the best features of dBB and MWI and avoiding their inherent problem. The new interpretation has been called the approach of Many Interacting Worlds (MIW).

\subsection{MIW: Let there be interaction!}\label{ssec:MIW}

In 2014 Hall, Deckert and Wiseman \cite{MIW} has proposed a novel and very interesting approach to quantum mechanics. Let's again consider a $D$-dimensional phase space, only this time we will introduce not just one world-particle, but $N$ of them, all cohabiting the same phase space, with $\vec x_k$ denoting the coordinate of the $k$-th world-particle. This set comprises different versions of the same world-particle -- in essence, we are dealing with a multiverse like we did in MWI. However, unlike MWI, let us next assume that these world-particles are {\em interacting with each other} via some sort of ``quantum interaction potential'' $U$, and that $U$ in the limit $N\to\infty$ turns into the well-known quantum potential $Q$, introduced in dBB. This way we not only dissipate the problem of particles behaviour predicated by some strange external force as it was in dBB, but also introduce a many worlds framework that does not depend on the wave function formalism, which now arises as a by-product at the limit $N\to \infty$.

Of course, in order for Many Interacting Worlds approach to actually work we must be able to construct the corresponding potential. The trick here lies in the assumption that for sufficiently large $N$ the standard quantum mechanical average of an observable function $\vphi$ must correspond with the average value on the set of all world-particles (here we use the fact that we cannot know which one of the world-particles is ours, and forced to stick to averages). In other words:
\beq \label{averages}
\left<\vphi\right> = \langle\psi|\vphi|\psi\rangle \approx \frac{1}{N} \sum\limits_{j=1}^N \vphi(\vec x_j),
\enq
which means that the distribution of $\vec x_j$ follows the probability density $|\psi|^2$ -- or, alternatively, that the modulus of wave function is a smoothed-over approximation of said distribution:
\beq \label{psi_delta}
|\psi|^2 \approx \frac{1}{N} \sum\limits_{n=1}^N \delta \left(\vec q - \vec x_n(0)\right).
\enq

For the sake of simplicity let us instead of $N$ world-particles (which are by definition amalgamations of multiple different particles) consider $N$ ordinary particles, mutually interacting with each other. Since all of them are versions of one particle (again, like in MWI), all of them have the same mass $m$. By assumption, our ensemble of particles must be controlled by a hamiltonian $H_N$:
\beq \label{H}
H_N = \sum\limits_{j=1}^N \frac{(\vec p_j)^2}{2m} + \sum\limits_{j=1}^N V(\vec x_j) + U_N(\vec x_1, \vec x_2,... \vec x_N),
\enq
where $\vec p_j$ is a momentum of $j$-th particle, $V(\vec x_j)$ are classic potentials, and $U_N$ is a new potential of quantum interaction. The average energy $\left< E \right>$ for our ensemble is simply
\beq\label{E_H}
\left< E \right> = \frac{H_N}{N}.
\enq
On the other hand, from the point of view of classical quantum mechanics the average energy is equal to
\beq \label{E}
\left< E \right> = \int dq ~|\psi|^2 \left(\frac{1}{2m}\left(\nabla S\right)^2 + V + \frac{\hbar^2}{8m} \left(\frac{\nabla P}{P}\right)^2\right),
\enq
where we have introduced a new useful variable $P$, which stand for the probability distribution $P=|\psi|^2=r^2$. Using \eqref{psi_delta}, we can rewrite \eqref{E} as:
\beqn
\left< E \right> \approx \frac{1}{N}\sum\limits_{j=1}^N \left(\frac{1}{2m}\left(\nabla S\right)^2 + V + \frac{\hbar^2}{8m} \left(\frac{\nabla P}{P}\right)^2\right)\Bigg|_{\vec q = \vec x_j}.
\enqn
Furthermore, we require that (at least for sufficiently large $N$) the MIW must be fully reconciled with dBB, so every ``doppelg\"anger'' particle from our set must satisfy the condition \eqref{p_S}:
\beq \label{p_j}
\vec p_j = m \frac{\partial\vec x_j}{\partial t} = \nabla S(\vec x_j),
\enq
this yielding the following:
\beq\label{E_N}
\left< E \right> \approx \frac{1}{N}\sum\limits_{j=1}^N \left(\frac{(\vec p_j)^2}{2m} + V(\vec x_j) + \frac{\hbar^2}{8m} \left(\frac{\nabla P(x_j)}{P(x_j)}\right)^2\right).
\enq
Finally, comparing \eqref{E_N} with \eqref{E_H} and \eqref{H}, we conclude that the sought after quantum interaction potential $U_N$ must have a general form:
\beq\label{U_N}
U_N \approx \sum\limits_{j=1}^N \frac{\hbar^2}{8 m} \left|\frac{\nabla P(x_j)}{P(x_j)}\right|^2.
\enq

This result can be easily illustrated if $N$ doppelg\"anger particles are restricted to a one-dimension phase space. First, let us assume that the coordinates $x_j$ of our particles are numerically ordered: $x_1<x_2<..<x_N$. Next, consider a slowly varying function $\phi$, defined on the entire phase space. We already know that its average may be estimated as:
\beq \label{phi}
\left< \phi \right> = \frac{1}{N} \sum \limits_{j=1}^N \phi(x_j) \approx \int dq P(q) \phi(q),
\enq
where the integration is supposed to go over the entire domain $q\in(-\infty,+\infty)$. On the other hand, thanks to \eqref{psi_delta}, we can restrict the domain of integration strictly to the area presently occupied by the $N$ particles. In fact, for sufficiently large $N$ and adequately slow variability of function $\phi$ we can introduce two auxiliary parameters $x_0=-\infty$, $x_{N+1}=-\infty$ and approximate the integral in \eqref{phi} as
\beq\label{Deriving_P}
\begin{split}
\frac{1}{N} \sum \limits_{j=1}^N \phi(x_j) &\approx \sum \limits_{j=1}^{N} \int \limits_{x_j}^{x_{j+1}} dq P(x_j) \phi(x_j) \\
&= \sum \limits_{j=1}^{N} (x_{j+1} - x_j) P(x_j) \phi(x_j).
\end{split}
\enq
For \eqref{Deriving_P} to hold for any combination of $x_j$'s, it must satisfy the following simple rule:
\beq \label{Px_j}
P(x_j) = \frac{1}{N(x_{j+1}-x_j)} \approx \frac{1}{N(x_{j-1}-x_j)}.
\enq
We are now ready to derive the explicit form of quantum potential $U_N$ from \eqref{U_N}. The scalar product of two gradients of $P(x_j)$ in one dimension turns into a squares of a derivative takens w.r.t. variables $x_j$.
If $N$ is large enough, the derivative can be approximated by:
\beqn
\frac{d P(x_j)}{d x_j} \approx \frac{P(x_{j+1}) - P(x_j)}{x_{j+1}-x_j},
\enqn
which, applied to \eqref{U_N} together with \eqref{Px_j}, produces the following formula:
\beq \label{UNN}
U_N = \frac{\hbar^2}{2m} \sum\limits_{j=1}^N \left(\frac{1}{x_{j+1}-x_j}-\frac{1}{x_j-x_{j-1}}\right)^2.
\enq
This potential can be shown to converge to the familiar dBB potential \eqref{Q_Bohm} \cite{MIW}. It is elegant in form yet repulsive by nature: it forces the ``doppelg\"anger'' particles to be repulsed from each other. This is the reason why the usual quantum phenomena are so fragile and fleeting: they are produced by high values of $U_N$, which occurs only when sufficiently many copies of the same particle are sufficiently close to each. But such a configuration is generally  unstable; once disturbed -- for example, during the observation, -- the particles will quickly scatter all over the phase space, dramatically increasing the distances between each other, reducing the quantum potential $U_N$ to near-zero values and, in the process, becoming classical themselves. However, this is where the things get really interesting. We already know that the in the framework of MIW we can speak not just of particles but also describe the whole collections of particles -- entire objects, manifested as ``world-particles''. The rules for them remain the same: a world-particle is thought of as a single specimen from a whole array of its copies, distributed on the general multi-dimensional phase space. Hence, unlike all the other interpretations, the MIW explicitly states: the quantum effects might be registered even for large, macroscopic objects, as long as there are at least some copies of said objects with near-similar parameters. If there are, then the quantum potential $U_N$ becomes large enough and the object experiences quantum interaction with its ``doppelg\"angers''. The only reason we don't see anything like that is due to the sheer amount of parameters that characterize the typical macroscopic objects: the dimension of their phase space is so large that it becomes virtually impossible to discover even two copies of the corresponding world-particles in a sufficiently close proximity. And yet there exists at least one special type of physical objects which are large enough yet require less than a handful of parameters for their descriptions. These objects are {\em Friedman universes} and we shall deal with them in the next section.

\subsection{MIU: A World is Not Enough}\label{ssec:MIU}
Consider a homogenous isotropic universe with a given curvature $\kappa$ (equal to either 0 or $\pm 1$), filled by matter fields of density $\rho$ and pressure $p$. Such a universe is known to be governed by a single dynamic parameter -- scale factor $a$, which satisfies the two Friedman-Lema\^itre-Robertson-Walker equations:
\beq\label{adot}
\frac{\dot a^2}{2} -\frac{4 \pi G}{3} a^2 \rho = - \frac{\kappa c^2}{2},
\enq
and
\beq\label{addot}
\ddot a = -\frac{1}{2}\left(\rho +\frac{3p}{c^2}\right) a,
\enq
whereas the density $\rho$ satisfies the continuity equation
\beq \label{rho}
\dot \rho = -3\left(\rho + \frac{p}{c^2}\right) \frac{\dot a}{a}.
\enq

As we have argued in the previous article \cite{YY}, the fact that a Friedman universe like this depends on a single parameter makes it uniquely susceptible to the quantum interaction in the MIW framework. Granted, the idea of multiple {\em universes} coexisting along with ours might be harder to swallow than accepting a cohort of multiple versions of a single particle -- or it would have been if we didn't already know of at least three different ``multiversal layers'' (see \cite{Tegmark03} and \cite{Tegmark07})!

Of course, in order to add the quantum interaction, we must first assume that there exists a set of $N$ universes with scale factors $a_n$, which are scale factor-ordered: $a_1<a_2<...<a_N$. Next, we have to pin down the Hamiltonian of a system of many interacting universes (MIU). For that end consider first a Hamiltonian of a single (non-interacting) universe. Fortunately, it is easy to see that the Hamiltonian for a Friedman universe is
\beqn
\mathcal{H}(a,\p) = \frac{\mathbf{p}^2}{2} - \frac{4 \pi G}{3} \rho(a) a^2 = -\frac{\kappa c^2}{2},
\enqn
where $\p=\dot a$ plays the role of a generalized momentum. Knowing $H$, we can use the inverse Legendre transform to produce the Lagrangian:
\beqn
L = \frac{1}{2} \dot a^2 + \frac{4 \pi G}{3} \rho(a) a^2,
\enqn
and it is then a short hop from here to the Friedman equations \eqref{adot} and \eqref{addot}, since both follows from the Euler-Lagrange equation coupled with a continuity condition \eqref{rho}. Knowing this simple technique we can now try to make a leap from a single universe to multiple interacting universes. We begin by assigning to each one of the interacting universes their own densities $\rho_n$, pressures $p_n$ and constant curvatures $\kappa_n$, all the while retaining the continuity equations (they are important because they ensure the conservation of energy):
\beq
\dot \rho_n = -3\left(\rho_n + \frac{p_n}{c^2}\right) \frac{\dot a_n}{a_n}.
\enq
The interaction between MIU should manifest itself as a new term in the upgraded Hamiltonian $\mathcal{H}_N$:
\beqn
\begin{split}
\mathcal{H}_N =& \sum\limits_{n=1}^N \frac{\mathbf{p}_n^2}{2} - \frac{4 \pi G}{3}\sum_{n=1}^{N}\rho_{n}(a_{n})a_{n}^{2} +\\
&+ U (a_1,..,a_{N})+\frac{c^2}{2}\sum_{n=1}^{N}\kappa_{n} = 0,
\end{split}
\enqn
where by analogue with the MIW framework we discussed in the previous section the quantum interaction potential should be of a form:
\beq
U(a_{1},..,a_{N}) = \alpha^{2}\sum_{n=1}^{N}\left(\frac{1}{a_{n+1}-a_{n}}-\frac{1}{a_{n}-a_{n-1}}\right)^{2},
\label{Tema000}
\enq
where $\alpha^2$ is a constant which can be shown to be equal to $\alpha^2=8\pi \hbar G/3c=\left(L_{_{PL}}c\right)^2$ \cite{YY}, and we have also introduced two auxiliary constants $a_0=a_{N+1}=\infty$. Following our algorithm, we can derive the modified Lagrangian
\beqn
\begin{split}
L_N &=\frac{1}{2}\sum_{n=1}^{N}\dot{a}_{n}^{2}+\frac{4 \pi G}{3}\sum_{n=1}^{N}\rho_{n}(a_{n})a_{n}^{2} -\\
&- \alpha^{2}\sum_{n=1}^{N}\left(\frac{1}{a_{n+1}-a_{n}}-\frac{1}{a_{n}-a_{n-1}}\right)^{2},
\end{split}
\enqn
and then plug it into the Euler-Lagrange equations
\beqn
\frac{d}{dt}\frac{\partial L}{\partial \dot{a}_{n}}=\frac{\partial L}{\partial a_{n}}, \qquad n = 1,..,N,
\enqn
which at last produces our brand new modified Friedmann equations:
\begin{equation}
\begin{split}
\ddot{a}_{n} &= -\frac{1}{2}(\rho_{n}+3p_{n})a_{n} -\\
&- \alpha^{2}\frac{\partial}{\partial a_{n}}\sum_{k=1}^{N}\left(\frac{1}{a_{k+1}-a_{k}}-\frac{1}{a_{k}-a_{k-1}}\right)^{2},
\end{split}
\label{5STAR}
\end{equation}
just as we required.
\end{appendices}
\newline
\section*{Acknowledgments} \label{Sec:Acknowledgements}
The article was supported by the Ministry of Science and Higher Education of the Russian Federation (agreement no. 075-02-2025-1789)
\newline
\newline


\begin{thebibliography}{99}


\bibitem{MIW} M. J. W. Hall, D.-A. Deckert and H. M. Wiseman, ''Quantum phenomena modelled by interactions between many classical worlds'', {\em Phys. Rev. X} {\bf 4} (2014) 041013.

\bibitem{YY} A. V. Yurov, V. A. Yurov,  ''The day the universes interacted: quantum cosmology without a wave function'', {\em Eur. Phys. J. C} {\bf 79}  (2019) 771.

\bibitem{Zero-Univ}  John D. Barrow, ''Is the universe ill-posed?'', http://arxiv.org/abs/2003.14108v2.

\bibitem{Susk-B}  Raphael Bousso and Leonard Susskind, ''Multiverse interpretation of quantum mechanics'', {\em Phys. Rev. D} {\bf 85} (2012) 045007.

\bibitem{Entropy1}  Roger Penrose, ''Cycles of Time: An Extraordinary New View of the Universe'',  Random House (USA) (2010).

\bibitem{Entropy2} Sean Carroll, ''From Eternity to Here: The Quest for the Ultimate Theory of Time'', Hardcover – January 7, (2010)
\bibitem{3.1} M. Bouhmadi-Lopez, P. F. Gonz\'{a}lez-D\'{i}az  and P. Martin-Moruno,''Worse than a big rip?'', {\em Phys. Lett. B} {\bf 659} (2008) 1--5.

\bibitem{3.2} A. V. Yurov, A. V. Astashenok and P.F. Gonz\'{a}lez-D\'{i}az, ''Astronomical bounds on future big freeze singularity'',  {\em Grav. Cosmol.} {\bf 14} No. 3  (2008) 205--212.

\bibitem{3.3} F. Cannata, A. Yu. Kamenshchik and D. Regoli, ''Scalar field cosmological models with finite scale factor singularities'',  {\em Phys. Lett. B} {\bf 670} (2009) 241--245.

\bibitem{3.4} Y. Shtanov and V. Sahni, ''New Cosmological Singularities in Braneworld Models'',  {\em Class. Quant. Grav.} {\bf 19} (2002) L101--L107.
\bibitem{5.1} J. D. Barrow, ''Sudden Future Singularities'',  {\em Class. Quant. Grav.} {\bf 21} (2004) L79--L82.

\bibitem{5.2} J. D. Barrow, ''More General Sudden Singularities'',   {\em Class. Quant. Grav.} {\bf 21} (2004) 5619--5622.


\bibitem{5.3} J. D. Barrow,''New Isotropic and Anisotropic Sudden Singularities'',  {\em Class. Quant. Grav.} {\bf 22} (2005) 1563--1571.

\bibitem{5.4} J. D. Barrow, ''New Anisotropic Sudden Singularities and Dimensional Reduction'',   {\em Phys. Rev. D} {\bf 102}  (2020) 024073.

 \bibitem{5.5} John D. Barrow, Spiros Cotsakis and Dimitrios Trachilis, ''The Generic Sudden Singularity in Brans-Dicke Theory'',  {\em Eur. Phys. J. C} {\bf 80}  (2020)  1197.
\bibitem{BBtS} A.O. Barvinsky, C. Deffayet and A. Y. Kamenshchik, ''Anomaly Driven Cosmology: Big Boost Scenario and AdS/CFT Correspondence'', {\em JCAP} 0805 (2008) 020.
\bibitem{BBS1} V. Gorini, A. Kamenshchik, U. Moschella and V. Pasquier,''Tachyons, Scalar Fields and Cosmology'',  {\em Phys. Rev. D} {\bf 69} (2004) 123512.

\bibitem{BBS2} Z. Keresztes, L. A. Gergely, V. Gorini, U. Moschella and A. Yu. Kamenshchik, ''Tachyon cosmology, supernovae data and the Big Brake singularity'', {\em Phys. Rev. D} {\bf 79} (2009) 083504.

\bibitem{NOT} S. Nojiri, S.D. Odintsov and S. Tsujikawa, ''Properties of singularities in (phantom) dark energy universe'',  {\em Phys. Rev.  D} {\bf 71}  (2005) 063004


\bibitem{SL}  A. V. Yurov, ''Brane-like singularities with no brane'',  {\em Phys. Lett. B} {\bf 689} (2010) 1--7.

\bibitem{JC} Artyom V. Yurov, Artyom V. Astashenok and Valerian A. Yurov, ''The Cosmological Models with Jump Discontinuities'',  {\em Eur. Phys. J. C} {\bf 78} (2018) 542.

\bibitem{JOPAQuanSing} A. Yu. Kamenshchik, ''Quantum cosmology and late-time singularities'', {\em Class. Quant. Grav.} {\bf 30} (2013) 173001.

\bibitem{JOPAVasilev} Teodor Borislavov Vasilev, Mariam Bouhmadi-López, Prado Martín-Moruno, ''Classical and Quantum $f(R)$ Cosmology: The Big Rip, the Little Rip and the Little Sibling of the Big Rip'',  Universe {\bf 7} (2021) 8, 288.

\bibitem{Astashenok} A private correspondence with Artyom V. Astashenok.



\bibitem{Linde2005} Andrei Linde, ''Particle Physics and Inflationary Cosmology'',  Contemp. Concepts Phys. 5 (2005) 1--362.

\bibitem{MuhChib1981} V. F. Mukhanov  and G. V. Chibisov, ''Quantum Fluctuation And `Nonsingular' Universe'', JETP Lett.  {\bf 33} (1981) 549--553   [Pisma Zh. Eksp. Teor. Fiz.  33, 549 (1981)].

\bibitem{Hawk1982} S. W. Hawking,   ''The Development Of Irregularities In A Single Bubble Inflationary Universe'', Phys. Lett. B 115 (1982) 295--297.

\bibitem{Star1982} A. A. Starobinsky,   ''Dynamics Of Phase Transition In The New Inflationary Universe Scenario And Generation Of Perturbations'', Phys. Lett. B 117 (1982) 175--178.


\bibitem{DeWitt1967} B. S. DeWitt, Phys. Rev. 160, 1113 (1967).
\bibitem{Wheeler1968} J. A. Wheeler, in: Relativity, Groups, and Topology, edited by C. M. DeWitt and J. A. Wheeler, Benjamin, New York (1968).




\bibitem{Hooft1993} G. ’t Hooft, ''Dimensional Reduction in Quantum Gravity'', ArXiv:gr-
qc/9310026.

\bibitem{Sussk1994}  L. Susskind, ''The World as a Hologram'',   Journal of Mathematical Physics, 36 (1995) 6377--6396.



\bibitem{Bek1972} J.D. Bekenstein, ''Black Holes And The Second Law of Thermodynamics'',
Lett. Nuovo Cim. 4 (1972) 737-740.












\bibitem{Vil1982} A Vilenkin, ''Creation of universes from nothing'', {\em Physics Letters B} {\bf 117} (1982) 25--28.
\bibitem{EQ} A.D. Chernin, D.I. Santiago, A.S. Silbergleit,  ''The interplay between gravity and quintessence: a set of new GR solutions'', {\em Physics Letters A} {\bf 294} (2002) 79--83.

\bibitem{Omega} Frank J. Tipler, Jessica Graber, Matthew McGinley, Joshua Nichols-Barrer, Christopher Staecker, ''Closed universes with black holes but no event horizons as a solution to the black hole information problem'', Monthly Notices of the Royal Astronomical Society, Volume 379, Issue 2, 1 August 2007, Pages 629–640.





\bibitem{MWI}  Hugh Everett ``The Theory of the Universal Wavefunction (1955)'': In Bryce DeWitt, R. Neill Graham (eds.). ``The Many-Worlds Interpretation of Quantum Mechanics'', {\em Princeton Series in Physics}, Princeton University Press (1973), 3--140.



\bibitem{Bohm} David Bohm, ``A Suggested Interpretation of the Quantum Theory in Terms of `Hidden' Variables (Parts I and II)'', {\em Phys. Rev.} {\bf 85} (1952) 170--193.



\bibitem{How} Stephen Hawking, George Ellis, ``The Large Scale Structure of Space-Time'', Cambridge University Press (1973).

\bibitem{2.1} R. R. Caldwell,   ''A Phantom Menace? Cosmological consequences of a dark energy component with super-negative equation of state'',  {\em Phys. Lett. B} {\bf 545} (2002) 23--29.
\bibitem{2.2} R. R. Caldwell, M. Kamionkowski and N. N. Weinberg, ''Phantom Energy and Cosmic Doomsday'', {\em Phys. Rev. Lett.} {\bf 91} (2003) 071301.
\bibitem{2.3} S. M. Carroll, M. Hoffman and M. Trodden, ''Can the dark energy equation-of-state parameter $w$ be less than -1?'',   {\em Phys. Rev. D} {\bf 68} (2003) 023509.

\bibitem{2.4} P. F. Gonz\'{a}lez-D\'{i}az, ''Achronal cosmic future'',  {\em Phys. Rev. Lett.} {\bf 93} (2004) 071301.

\bibitem{2.5} A. V. Yurov, P. M. Moruno and P. F. Gonz\'{a}lez-D\'{i}az, ''New ''Bigs'' in Cosmology'',  {\em Nucl. Phys. B} {\bf 759} (2006) 320--341.








\bibitem{PedroCarmen}  Pedro F. Gonz\'{a}lez-D\'{i}az, Carmen L. Siguenza, ''Phantom thermodynamics'',  {\em Nucl. Phys. B} {\bf 697} (2004) 363--386.

\bibitem{LimaAlcaniz}  J.A.S. Lima, J.S. Alcaniz, ''Thermodynamics, spectral distribution and the nature of dark energy'',
{\em Phys. Lett. B} {\bf 600} (2004) 191--196.

\bibitem{BRRaz}  Artyom Yurov, ''Phantom scalar fields result in inflation rather than Big Rip'', arXiv:astro-ph/0305019; Yurov, A.V. ''Phantom scalar fields result in inflation rather than Big Rip'', {\em Eur. Phys. J. Plus} {\bf 126}  (2011) 132.


\bibitem{Neopr}  A. V. Yurov, V. A. Astashenok, V. A. Yurov, ''The dressing procedure for the cosmological equations and the indefinite future of the universe'',  {\em Grav. Cosmol.} {\bf 14} (2008) 8--16.



\bibitem{VilGar2001} Jaume Garriga and Alexander Vilenkin, ''Many worlds in one'', {\em Phys. Rev. D} {\bf 64} (2001) 043511.

\bibitem{LinVan2010}  Andrei Linde, Vitaly Vanchurin, ''How many universes are in the multiverse?'', {\em Phys. Rev. D} {\bf 81} (2010) 083525.

\bibitem{Kas} Edward Kasner, ''Geometrical Theorems on Einstein's Cosmological Equations'', {\em American Journal of Mathematics}
{\bf 43} No. 4 (1921) 217--221.

\bibitem{Bel} V. A. Belinskii, I.M. Khalatnikov, ''Effect of Scalar and Vector Fields on the Nature of the Cosmological Singularity'',  {\em Sov. Phys. JETP} {\bf 36}  (1973) 591.

\bibitem{Kamen} A. Yu. Kamenshchik,  ''The problem of singularities and chaos in cosmology'',  {\em Phys. Usp.} {\bf 53}  (2010) 301--309.



\bibitem{Durr} D. D\"urr, S. Goldstein and N. Zangh\'i, ``Quantum equilibrium and the origin of absolute uncertainty''. {\em J. Stat. Phys.} {\bf 67}, 843--907 (1992).

\bibitem{Holland} Peter R. Holland, ``The Quantum Theory of Motion: An Account of the De Broglie-Bohm Causal Interpretation of Quantum Mechanics'', Cambridge University Press, Cambridge (1993).

\bibitem{Licata} Ignazio Licata \and Davide Fiscaletti, ``Quantum potential: Physics, Geometry and Algebra'', AMC, Springer (2013).

\bibitem{Tegmark03} Max Tegmark, ``Parallel Universes'', Science and Ultimate Reality: From Quantum to Cosmos (honoring John Wheeler's 90th birthday), ed. J. D. Barrow, P.C.W. Davies \and C.L. Harper. Cambridge University Press (2003)

\bibitem{Tegmark07} Max Tegmark, ``The Multiverse Hierarchy'', Universe or Multiverse?, ed. B. Carr, Cambridge University Press (2007)





\end{thebibliography}
\end{document}